\documentclass[a4paper,fleqn]{cas-sc}

\usepackage[authoryear,longnamesfirst]{natbib}
\usepackage{amssymb}
\usepackage{amsmath}
\usepackage{amsmath,amsfonts}
\usepackage{algorithmic}
\usepackage{algorithm}
\usepackage{array}
\usepackage{caption}
\usepackage{subfigure}
\usepackage{textcomp,color}
\usepackage{stfloats}
\usepackage{url}
\usepackage{verbatim}
\usepackage{graphicx}
\usepackage{cite}
\usepackage{amsthm}
\usepackage{hyperref}
\newtheorem{theorem}{Theorem}
\theoremstyle{definition}
\newtheorem{definition}{Definition}
\newtheorem{lemma}{Lemma}
\newtheorem{assumption}{Assumption}
\newtheorem{remark}{Remark}

\newcommand{\remove}[1]{}

\DeclareMathOperator*{\argmax}{arg\,max}
\def\tsc#1{\csdef{#1}{\textsc{\lowercase{#1}}\xspace}}
\tsc{WGM}
\tsc{QE}
\tsc{EP}
\tsc{PMS}
\tsc{BEC}
\tsc{DE}

\begin{document}
\let\WriteBookmarks\relax
\def\floatpagepagefraction{1}
\def\textpagefraction{.001}
\shorttitle{Content and Access Networks Synergies}
\shortauthors{P. Agarwal et~al.}

\title [mode = title]{Content and Access Networks Synergies: Tradeoffs in Public and Private Investments by Content Providers}                      
\tnotemark[1]

\tnotetext[1]{Part of this work was done when Pranay Agarwal was an Institute Postdoctoral Fellow in the Department of Electrical Engineering at IIT Bombay.}


\author[1]{Pranay Agarwal}[
                        orcid=0000-0003-4323-7051]
\cormark[1]

\credit{Part of this work was done when I was an Institute Postdoctoral Fellow in the Department of Electrical Engineering at IIT Bombay.}

\affiliation[1]{organization={Department of Electrical and Electronics Engineering, BITS Pilani Hyderabad Campus},
                city={Hyderabad},
                postcode={500078}, 
                state={Telangana},
                country={India}}

\author[2]{D. Manjunath}

\ead{dmanju@ee.iitb.ac.in}


\affiliation[2]{organization={Department of Electrical Engineering, IIT Bombay},
                postcode={400076}, 
                city={Mumbai},
                state={Maharashtra},
                country={India}}

\cortext[cor1]{Corresponding author}


\begin{abstract}
The ubiquity of smartphones has fueled content consumption worldwide, leading to an ever-increasing demand for a better Internet experience. This has necessitated an upgrade of the capacity of the access network. The Internet service providers (ISPs) have been demanding that the content providers (CPs) share the cost of upgrading access network infrastructure. A \emph{public investment} in the infrastructure of a neutral ISP will boost the profit of the CPs, and hence, seems a rational strategy. A CP can also make a \emph{private investment} in its infrastructure and boost its profits. In this paper, we study the trade-off between public and private investments by a CP when the decision is made under different types of interaction between them. Specifically, we consider four interaction models between CPs---centralized allocation, cooperative game, non-cooperative game, and a bargaining game---and determine the public and private investment for each model. Via numerical results, we evaluate the impact of different incentive structures on the utility of the CPs. We see that the bargaining game can result in higher public investment than the non-cooperative and centralized models. However, this benefit gets reduced if the CPs are incentivized to invest in private infrastructure.
\end{abstract}

%


\maketitle

\section{Introduction}

\begin{figure}
    \centering
    \includegraphics[width=3.5in]{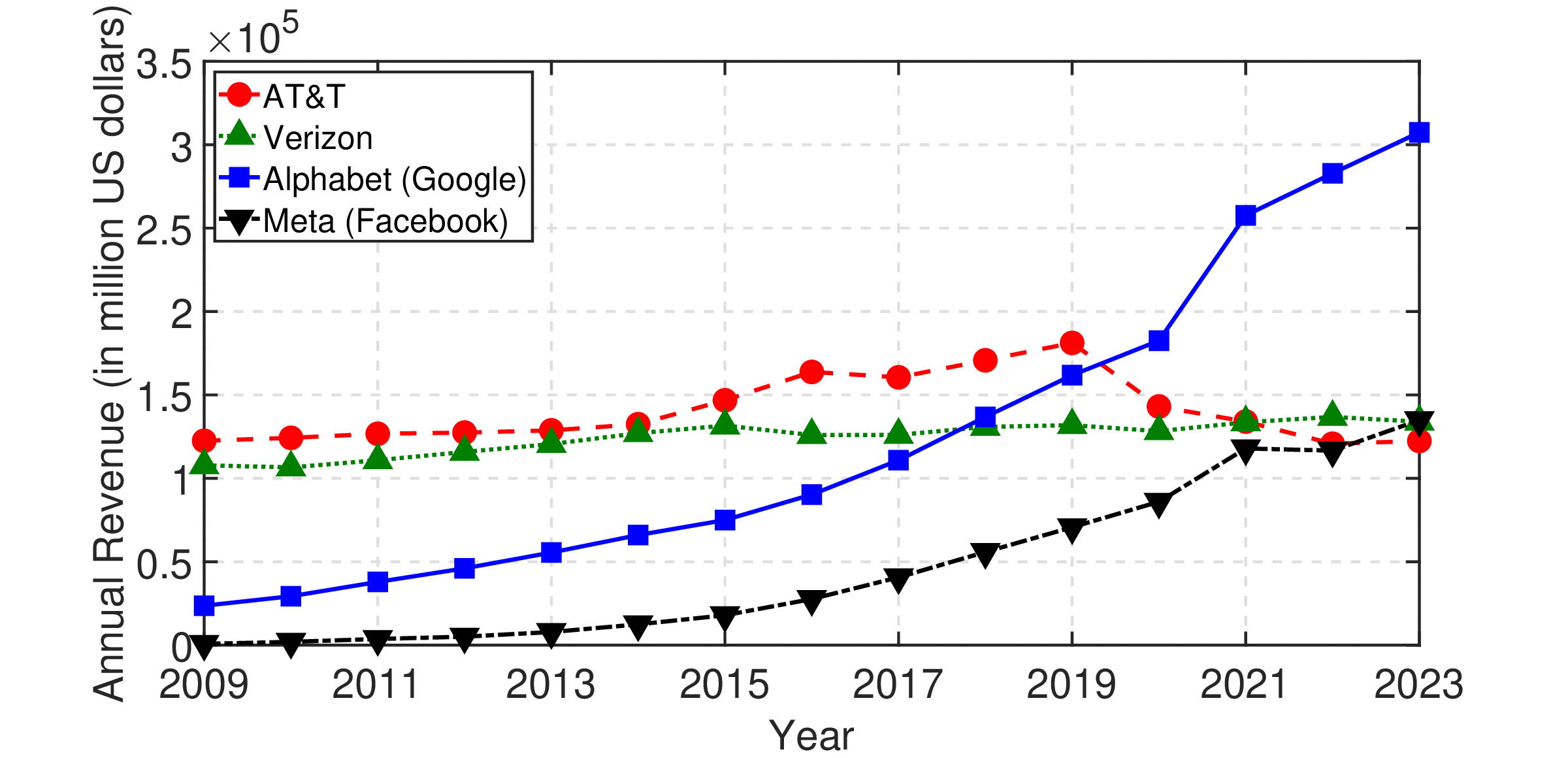}
    \caption{Annual revenue (in million US dollars) of AT\&T, Verizon, Alphabet, and Meta.}
    \label{fig:revenue}
\end{figure}

Internet service is provided to users by two complementary networks---access provider networks (APs) and content provider (CP) networks. Most content providers use specialized worldwide content networks, which could be their own or they could contract with content distribution networks (CDNs) that provide such a service. Clearly, APs and CPs are complementary services; without one, the other is significantly less valuable. Despite the complementarity of the two services, contrary to the conventional wisdom of classical economic theory, CPs have been able to monetize their resources significantly more effectively than the APs; GSMA estimates the connectivity industry to be 14\% of the total value of the Internet chain in 2022 \citet{Kearney22} and this fraction has been steadily decreasing from about 21\% in 2008. Perhaps a starker comparison could be between the annual revenues of two large APs (AT\&T and Verizon) and two large content providers (Alphabet and Meta) over the last twenty years (see Fig. \ref{fig:revenue}); APs have stayed flat while those of CPs have been growing. 

While the share of the Internet Value Chain pie for the APs has been decreasing steadily, technological advances in content creation and rendering have made steady advances. The burden of delivering these advances is with the APs, which in turn have to make heavy investments, while the CPs are expected to reap a larger share of the economic surplus that will result from this investment. Because of this, there is an argument, at least from some of the ISPs, that the CPs must bear a share of such investments in the access networks\footnote{https://www.telefonica.com/en/communication-room/blog/economic-analysis-gives-evidence-for-market-failure-contributions-from-big-tech-needed-to-unlock-consumer-welfare}. Alternatively, since the APs do not always make the necessary investments (or cannot afford to make them), the CPs can invest to improve access. And indeed, CPs have been making several such investments. For example, Google has been providing free WiFi in many public places\footnote{https://indianexpress.com/article/technology/social/google-station-free-wifi-moves-to-smart-cities-150-hotspots-to-go-live-in-pune-5046151/}. They have also invested in municipal WiFi projects and rolled out GoogleFi. Facebook, before it became Meta, had also invested in providing access, e.g., through the FreeBasics\footnote{See, e.g.,https://www.facebook.com/connectivity/solutions/free-basics\\ and also https://www.theguardian.com/technology/2016/may/12/facebook-free-basics-india-zuckerberg} \footnote{The walled garden nature of FreeBasics was a source of concern, and it was banned in many markets.}. A study of the quality of the service of Freebasics is reported in \citet{Sen16}.

The investments in access mentioned above are not very restrictive to other content providers and can, in an idealized abstraction, be considered to be investments by the respective CPs in a public good. There is also another type of investment that the CPs are making to improve the delivery. They are building their own content distribution networks with significant caching capacity nearer the consumption points, e.g., Google, Meta, and Netflix. Many CPs are also buying such a service from third-party content distribution networks (CDNs) like Akamai and Lightstream; notably, CDNs also peer with access networks. This latter kind of investment can be termed as a private investment by a CP to improve its content delivery; the economics of such private connections between the networks of CPs and APs are studied in \citet{Patchala21}.

Thus, we have the following question to explore in this paper: Assume that the CPs have to pay for the infrastructure of a neutral ISP. Further, they also have an option of private investment, improving the quality of service (QoS) of their customers.
In this situation, what are the effects of the incentive structures and the interaction models among the CPs on the (i)~investments in the access networks, and on the (ii)~social utility. Our approach is to use an abstraction that models the impact of the private and public investments by the CPs on their revenue and analyzes the total investments made under various interaction models among CPs. This work has similarities with that in \citet{Kalvit2019} where only public investments of CPs were considered. 



\subsection{Related Literature}
Smart pricing techniques to potentially make the CPs and customers pay the `market price' for use of the access network have been typically considered under two variations; providing differentiated quality of service e.g., \citet{Choi10,Economides12a,Economides12b} or through differentiated pricing schemes like sponsored data and zero rating, e.g., \citet{Schewick14,Rossini15,Phalak19,Vyavahare22}. However, through a combination of market forces and legal interventions, net neutrality is either \textit{de jure} or \textit{de facto} in almost all markets. Hence, these smart pricing schemes are not really applicable in practice; they are also not directly germane to this work. A closely related work is \citet{Kalvit2019}, where they consider a voluntary payment by the CPs to the neutral ISP, which is to be used as an investment in the access network to improve the latter. In \citet{Kalvit2019}, the voluntary payment is treated like a game by the CPs to provide the common good (see, e.g., \citet{Varian10} for a definition) and the total contribution is analyzed under different types of games. In the same vein as \citet{Kalvit2019}, \citet{Bouvard22} proposes a profit-sharing model between the ISP and CPs. None of these considers the impact of private investments that the CPs are making. 


\subsection{Preview of our results}
This work considers a system of multiple CPs and a neutral ISP that together serve a fixed user population. An upgrade in the ISP access network boosts the QoS experienced by the consumers. This leads to enhanced consumer engagement, thereby increasing the volume of content consumed and consequently the profit of a CP. Thus, voluntary contribution towards expanding the capacity of an ISP infrastructure by a CP is to its benefit. Since the ISP is neutral, such an investment by a CP is for the public good because the upgraded ISP infrastructure enhances the QoS of the consumers of all the CPs. Hence, we term this as \emph{public investment.}  Alternatively, or perhaps in addition, a CP can also invest to improve its content delivery mechanisms, for example, by employing better caching facilities, improving encoding techniques, etc. Such an investment by a CP only caters to its consumers and is hence termed a \emph{private investment.} Thus, given the two components of a CP's investment---public and private---we aim to study the investment strategy of CPs in different settings. Specifically, we consider four models of interactions between the CPs, namely a centralized allocation, a strategic cooperative game, a non-cooperative game, and a bargaining game, and determine the public and private investments of the CPs for each model. 

The key findings of this work are as follows. We observe from numerical results that a stable grand coalition of all CPs is feasible in some scenarios. Unless symmetry exists between CPs, at most one CP contributes to the public investment in the non-cooperative game. However, all CPs make private investments. The bargaining model results in higher total public investment than the centralized game. However, this gain is reduced if the CPs have an incentive for private investment.

\subsection{Paper Organization}
The rest of this paper is organized as follows.
We first describe our model in Section \ref{sec:Model}, wherein we characterize the increase in the utility of a CP from the net public investment and its private investment and define the public-private trade-off, a central metric of interest in this work. Then, Section \ref{sec:Centralized} presents the centralized allocation model wherein CPs collectively aim to maximize their net utility. We discuss the strategic cooperative game in Section \ref{sec:cooperative}, wherein we define the notion of stability of the grand coalition of all CPs and determine the public and private investments of the CPs. Section \ref{sec:Nash} describes the non-cooperative game between the CPs wherein each CP aims to maximize its utility. The bargaining game between the CPs is discussed in Section \ref{sec:Bargaining}. In Section \ref{sec:Numericals}, we present and discuss some numerical results. Finally, Section \ref{sec:Conclusions} briefly summarizes the work done. All the proofs are provided in the Appendix.

\section{System Model}\label{sec:Model}
There are $N$ CPs and one ISP that together provide Internet services to a fixed user population
\footnote{In general, there will be multiple ISPs. Here, we consider the stylized model with one ISP to build intuition.}.
There is a basic infrastructure that is present at the ISP and the CPs, and the interest here is in new investments that need to be made to improve the quality of service for the users. Each CP can make a \emph{public investment} towards the improvement of the infrastructure of the neutral ISP; users of all the CPs see an improvement in the service quality from this investment. A CP can also invest in its private infrastructure to improve the service quality to only its users, and this is the \emph{private investment.} Let $q_n$ be the public investment and $p_n$ be the private investment of CP $n,$ for all $n \in \mathcal{N}$, where $\mathcal{N}:=\{1, 2, \ldots, N\}$. Let $Q := \sum_{n=1}^N q_{n}$ be the total public investment and the $P := \sum_{n=1}^N p_n$ be the total private investment. Our objective in this paper is to evaluate the relation between $P$ and $Q$ under different assumptions of interaction between the CPs. Specifically, we will be interested in, among others, the  measure
\begin{equation}
    \gamma := \frac{Q}{P} \label{eq:gamma} 
\end{equation}
that we call the public-private trade-off. 

The public and private investments lead to an improvement in the overall QoS experience of the users, a consequent increase in traffic, and hence, a revenue gain to the CPs. Further, since the ISP operates in the net neutrality regime, the gain is a function of the total public investment. A total public investment of $Q$ and a private investment of $p_n$ results in an increased consumption of the content of CP $n$ by an amount $g_n(Q+h_{n}(p_{n})),$ where both $g_n(\cdot)$ and $h_n(\cdot)$ are concave increasing functions. This is a reasonable assumption because the marginal benefit from increasing investments will decrease. 
Further, $h_{n}(\cdot)$ is such that the inequality 
\begin{equation}
    h_{n}(p_{n}) \geq p_{n} \label{eq:hn_cond}
\end{equation}
holds for some $p_{n}$ implying 
$$g_{n}(Q_{-n} + q_{n} + h(p_{n})) \geq g_{n}(Q_{-n} + q_{n} + p_{n})$$
for some $p_{n}$ and $q_{n}.$ Here $Q_{-n} := \sum_{m=1, m \neq n}^{N} q_{m}.$
This requirement only means that there exists some $(p_n, q_n)$ for which a CP will make non-zero public and private investments, a reasonable technical assumption. 
 
\begin{remark}
A possible view of $h_n(\cdot)$ is that it is the `equivalent public investment' for a private investment of $p_n.$ 
\end{remark}

Denoting the revenue per unit traffic for CP $n$ to be $r_n,$ the surplus, or the utility function for CP $n,$ denoted by $U_{n},$ is  
\begin{equation}
    U_{n}(q_{n}, p_{n}) = r_{n}g_n(Q+h_n(p_{n})) - (p_{n} + q_{n}) , \forall n \in \mathcal{N} . \label{eq:Un1}
\end{equation}

\begin{remark}
    Since $g_n(\cdot)$ and $h_n(\cdot)$ are concave increasing, it is easy to show that $U_n$ is concave in both $p_n$ and $q_n.$ This is not surprising because the marginal revenue gains from either the public or the private investment should eventually plateau, and the net gain will become negative. 
\end{remark}  

In the following sections, we will determine the $p_n$ and $q_n,$ $n \in \mathcal{N}$, when the CPs make their choices under four interaction models---a centralized allocation, a cooperative game, a non-cooperative game, and a bargaining game. In the process, we also obtain the $\gamma$ and the utility functions of the different players under these interactions. In the rest of the paper, we use the following forms of $g_n(\cdot)$ and $h_n(\cdot)$
\begin{equation}
\label{assm:g_h} 
g_n(x) = a_n \log(1+x) \text{ and } h_n(x)= b_n\sqrt{x}\, \,\,\,\,\, \forall \ \ x \geq 0 .
\end{equation}
Further, we impose that $b_{n} \geq 1$ to meet the structural requirement discussed in (\ref{eq:hn_cond}).
Substituting (\ref{assm:g_h}) into (\ref{eq:Un1}), we get
\begin{equation}
    U_{n}(q_{n}, p_{n}) = r_{n}a_{n} \log(1+Q+b_n \sqrt{p_{n}}) - (p_{n}+q_{n}) \, . 
    \label{eq:Un}
\end{equation}
\section{Centralized Allocation}
\label{sec:Centralized}

First, let us consider a centralised allocation where both the $Q$ and the $p_n$ are determined centrally to maximize the sum utilities of all the CPs  
\begin{equation}
    U_{C}(Q, \{p_n\}) := \sum_{n=1}^{N} \left( r_{n}a_{n} \log(1+Q+b_{n}\sqrt{p_{n}}) - p_{n} \right) - Q.   \label{eq:utility}
\end{equation}
This is an idealisation and will be used to get the best case scenario that will, in turn, be used as a standard to compare against. The optimization problem in this \emph{centralized} allocation will be 
\begin{align}
    &\max_{Q, \{p_{n}\}} \ \ \sum_{n=1}^{N} (r_{n}a_{n}\log(1+Q+b_{n}\sqrt{p_{n}}) - p_{n}) - Q , \nonumber \\
   \text{s. t. } & p_{n} \geq 0 ~\forall n \in \{1, \ldots, N\} , \nonumber \\
   & Q \geq 0 . \label{eq:centralized_problem}
\end{align}
Towards solving \eqref{eq:centralized_problem}, we first present the following lemma that characterizes the optimum $p_n$ for a given $Q.$  
\begin{lemma}\label{lem:opti_pn}
    The optimum value of $p_{n}$ for a given value of $Q$, denoted by $p_{n}^{\ast}(Q)$, is given by
    \begin{equation}
        p_{n}^{\ast}(Q) = \left(\frac{\sqrt{(1+Q)^{2}+2b_{n}^{2}r_{n}a_{n}}-(1+Q)}{2b_{n}}\right)^{2} . \label{eq:opti_pn}
    \end{equation}
\end{lemma}

\begin{remark}
Lemma \ref{lem:opti_pn} says that $p_{n}^{\ast}(Q)$ is strictly positive for all $Q \geq 0$ and is a convex decreasing function of $Q$.
\end{remark}
Thus, given a fixed total public investment, a CP can always increase its utility by making private investment, however small. Further, because of the concavity of the utility function, the marginal investment that can improve its utility decreases with increasing $Q.$   

The optimum total public investment with \emph{centralized allocation}, i.e., the solution to \eqref{eq:centralized_problem}, denoted by $Q_{C}^{\ast}$, is characterised by the following lemma.

\begin{lemma}\label{lem:lemma_coop}
$Q_{C}^{\ast}$ is obtained by solving 
\begin{equation}
    \sum_{n=1}^{N} \frac{\sqrt{(1+Q)^{2}+2b_{n}^{2}r_{n}a_{n}}}{b_{n}^{2}} = (1+Q)\left(\sum_{n=1}^{N} \frac{1}{b_{n}^{2}}\right) + 1. \label{eq:lemma_coop}
\end{equation}
\end{lemma}
Lemma \ref{lem:lemma_coop} implies 
\begin{equation}
    Q_{C}^{\ast}>0 \iff \sum_{n=1}^{N}\frac{\sqrt{1+2b_{n}^{2}r_{n}a_{n}}}{b_{n}^{2}} > 1 + \sum_{n=1}^{N} \frac{1}{b_{n}^{2}}. \label{eq:non-zero_cond}
\end{equation}

Denoting the public-private trade-off under \emph{centralized allocation} by $\gamma_{C},$ we have the following theorem. 
\begin{theorem} \label{thm:coop_gamma}
    For {centralized} allocation of $\{p_n\}$ and $Q,$  
    \begin{equation}
        \gamma_{C} = \frac{2Q_{C}^{\ast}}{\sum_{n=1}^{N}r_{n}a_{n} - (1+Q_{C}^{\ast})} . \label{eq:lemma_coop_gamma}
    \end{equation}
\end{theorem}
\section{Strategic Cooperative Game} \label{sec:cooperative}
In this subsection, we discuss the strategic cooperative form of interaction between CPs and ISP. It takes into consideration the rationality of the CPs, and hence, differs from the idealized centralized form of interaction.
In the strategic cooperative game, we first aim to define the grand coalition and discuss its stability. Then, we aim to determine the public and private investments of the CPs.

Let $\mathcal{I}$ denote the set of CPs that make a non-zero public investment. Specifically,
\begin{equation}
    \mathcal{I} := \{n \in \mathcal{N}: q_{n}>0\} . \nonumber
\end{equation}
Note that $\mathcal{I} \subseteq \mathcal{N}$. We denote the total public investment for the set $\mathcal{I}$ by $Q(\mathcal{I})$, and  $$Q(\mathcal{I}) := \sum_{n \in \mathcal{I}} q_{n}.$$
Since we consider a neutral ISP, the upgraded infrastructure benefits all the CPs equally. Consequently, the CPs can potentially be segregated as \emph{investing} and \emph{free-riding} CPs. The investing CPs constitute the set $\mathcal{I}$, i.e., they contribute towards the public infrastructure. Whereas, the free-riding CPs just reap the benefits of the total contribution made by the investing CPs. When $\mathcal{I} = \mathcal{N}$, i.e., all the CPs make a non-zero public investment, we term this coalition as the \emph{grand coalition}. In other words, in \emph{grand coalition}, there are no free-riding CPs. However, given the scope of free-riding, the grand coalition of all CPs is stable only if no CP or a group of CPs has any incentive to free-ride. The following definition formally describes this understanding of the stability of the grand coalition.
\begin{definition}[Stability of Grand Coalition] \label{def:grand_coalition}
Let $\mathcal{I}^{\ast}$ denote the special case of grand coalition when all the CPs make non-zero public investment to the ISP. This implies $\mathcal{I}^{\ast}=\mathcal{N}$. 
   Then, the grand coalition is stable if and only if $$U_{n}( Q(\mathcal{I}^{\ast}), p_{n}) > U_{n}(Q(\mathcal{I}), p_{n})$$ holds for all $n \in \mathcal{N}$ and all possible $\mathcal{I} \subset \mathcal{N}$. 
\end{definition}
Given Definition \ref{def:grand_coalition} and using (\ref{eq:Un}), the strategic cooperative form of interaction between CPs and the ISP is formulated as 
\begin{align}
    &\max_{\{q_{n}\}, \{p_{n}\}} \sum_{n=1}^{N} U_{n}(q_{n}, p_{n}) \nonumber \\
    \text{s. t. } &U_{n}(Q(\mathcal{I}^{\ast}), p_{n}) > U_{n}(Q(\mathcal{I}), p_{n}) , ~\forall n \in \mathcal{N}, ~\forall \mathcal{I} \subset \mathcal{N}, \nonumber \\
    & q_{n} > 0 , ~\forall n \in \mathcal{N}, \nonumber \\ 
    & p_{n} \geq 0 , ~\forall n \in \mathcal{N}. \label{eq:coalition_problem}
\end{align}
Specifically in (\ref{eq:coalition_problem}), we aim to determine a tuple of non-zero public investments of all the CPs, i.e., $\{q_{1}, q_{2}, \ldots, q_{N}\}$, which maximizes the net utility of all the CPs while ensuring the stability of grand coalition as specified in the Definition \ref{def:grand_coalition}. If there exists a solution of (\ref{eq:coalition_problem}), the core of the coalition game is said to be non-empty. It is analytically intractable to prove the non-emptiness of the core for any $N$. However, we will evaluate the same via numerical results in Section \ref{sec:Numericals}.

We observe from (\ref{eq:coalition_problem}) that the number of conditions required to ensure the stability of the grand coalition scale exponentially in $N$ as there are $2^{N}$ possible choices of $\mathcal{I}$. 
We now present the following assumptions on $r_{n}$, $a_{n}$, and $b_{n}$ that help us omit some choices of $\mathcal{I}$.
\begin{assumption}\label{assm:coalition}
    The $r_{n}$, $a_{n}$, and $b_{n}$ for all $n \in \mathcal{N}$ follow the following order: 
    $$r_{1}a_{1} > r_{2}a_{2} > \cdots > r_{N}a_{N} \text{ and } b_{1} \leq b_{2} \leq \cdots \leq b_{N}.$$
\end{assumption}
Under Assumption \ref{assm:coalition}, the following lemma presents the structural insights for some choices of $\mathcal{I}$.
\begin{lemma}\label{lem:coalition}
    Under Assumption \ref{assm:coalition}, if $\mathcal{I}=\{n\}$ for any $n \in \mathcal{N}$ such that $n>1$, the CP $m$ has the incentive to make a non-zero public investment, where $m \in \mathcal{N}$ and $m<n$.
\end{lemma}
A key takeaway from Lemma $\ref{lem:coalition}$ is that we need not consider $\mathcal{I}=\{n\}$ for $n \in \mathcal{N}$ such that $n>1$ while solving (\ref{eq:coalition_problem}) alleviating the computational complexity of solving (\ref{eq:coalition_problem}). For example, for $N=2$, we only consider $\mathcal{I}=\{\emptyset, \{1\}\}$ while solving (\ref{eq:coalition_problem}).

\section{Non-Cooperative Game}
\label{sec:Nash}
We now consider the case when the CPs choose the $p_n$ and $q_n$ strategically, i.e., they do not coordinate and choose these quantities to selfishly maximize their utility. Using Lemma \ref{lem:opti_pn}, we begin the analysis by first rewriting $U_{n}$ as 
\begin{equation}
U_{n}  =  r_{n}a_{n}f_{n}(Q) - p_{n}^{\ast}(Q) - q_{n} 
    \label{eq:Un_2}
\end{equation}
where
\begin{equation}
f_{n}(Q)  =  \log\left(\frac{\sqrt{(1+Q)^{2}\!+\!2b_{n}^{2}r_{n}a_{n}}\!+\!(1\!+\!Q)}{2}\right) . \label{eq:fnQ}
\end{equation}

Since $U_{n}$ is a function of $Q=\sum_n q_n, $ and the latter are the actions of the CPs, we determine the {Nash equilibrium} choice of $\{q_n\}$ as follows. Let $q_{n}^{\ast}$ denote the equilibrium public investment maximizing the utility of CP $n$. Then, a {pure strategy Nash equilibrium} exists if and only if 
\begin{equation}
    U_{n}(q_{n}^{\ast}, Q_{-n}^{\ast}) \geq U_{n}(q_{n},Q_{-n}^{\ast})
\end{equation}
holds for $\{q_{n}\}_{n \in \mathcal{N}}$ where $Q_{-n}^{\ast}=\sum_{m \in \mathcal{N}, m \neq n} q_{m}^{\ast}.$ In other words, $\{q_{n}^{\ast}\}_{n \in \mathcal{N}}$ is a {Nash equilibrium} public investment tuple that ensures
\begin{equation}
    q_{n}^{\ast} = \argmax_{q_{n}} \, U_{n}(q_{n},Q_{-n}^{\ast})~\forall n \in \mathcal{N}. 
\end{equation}
Now, using $Q_{N}^{\ast} := \sum_{n \in \mathcal{N}} q_{n}^{\ast},$ we present the following theorem.  
\begin{theorem} \label{thm:NC}
    For {non-cooperative} interaction, the following holds.
    \begin{itemize}
        \item If $\max \limits_{n \in \mathcal{N}} \left(r_{n}a_{n} - \frac{b_{n}^{2}}{2}\right) \leq 1$, then $q_{n}^{\ast}=0$ for all $n \in \mathcal{N}$, and hence, $Q_{N}^{\ast}=0$.
        \item If $\max \limits_{n \in \mathcal{N}} \left(r_{n}a_{n} - \frac{b_{n}^{2}}{2}\right)>1$, then let 
        \begin{equation}
            \mathcal{M} := \left\{m \in \mathcal{N}: m = \argmax\limits_{n \in \mathcal{N}}\left(r_{n}a_{n}-\frac{b_{n}^{2}}{2}\right)\right \}. \label{eq:set_M}
        \end{equation}
        Then, $q_{n}^{\ast} \geq 0$ for all $n \in \mathcal{M}$ and $q_{n}^{\ast}=0$ for all $n \in \mathcal{N} \setminus \mathcal{M}$. Further, $1+Q_{N}^{\ast} = \max \limits_{n \in \mathcal{N}} \left(r_{n}a_{n} - \frac{b_{n}^{2}}{2}\right)$.
    \end{itemize}
\end{theorem}

An important conclusion from Theorem \ref{thm:NC} is that only the CP which has the maximum value of $r_{n}a_{n}-\frac{b_{n}^{2}}{2}$ contributes to the public investment and the remaining CPs free-ride on this. If more than one CP has the maximum value of $r_{n}a_{n}-\frac{b_{n}^{2}}{2}$, i.e., some symmetry exists between the CPs, $q_n >0$ for all such CPs. All the CPs though invest in their private infrastructure and the optimum value of $p_{n}$ can be obtained by substituting $Q$ by $Q_{N}^{\ast}$ in \eqref{eq:opti_pn}. 

Let $\eta$ denote the price of anarchy defined as
\begin{equation}
    \eta := \frac{Q_{C}^{\ast}}{Q_{N}^{\ast}} . \label{eq:eta}
\end{equation}
Further, let $\gamma_{N}$ denote the public-private trade-off for the {non-cooperative} interaction. The following theorem provides insights on $\eta$ and $\gamma_{N}.$
\begin{theorem}\label{thm:PoA}
    For \emph{Non-Cooperative} interaction, the following holds.
    \begin{itemize}
        \item If $\max\limits_{n \in \mathcal{N}} \left(r_{n}a_{n}-\frac{b_{n}^{2}}{2}\right) \leq 1$ and $\sum_{n=1}^{N} \frac{\sqrt{1+2b_{n}^{2}r_{n}a_{n}}-1}{b_{n}^{2}} > 1$, then $\eta$ is unbounded and $\gamma_{N}=0$.
        \item If $\max\limits_{n \in \mathcal{N}} \left(r_{n}a_{n} - \frac{b_{n}^{2}}{2}\right) > 1$, then $\eta > 1$. For the special case of $|\mathcal{M}|=|\mathcal{N}|$, 
        \begin{equation}
            \gamma_{N} = \frac{Q_{N}^{\ast}}{\sum \limits_{n \in \mathcal{N}} \frac{b_{n}^{2}}{4}} . \label{eq:gamma_n}
        \end{equation}
    \end{itemize}
\end{theorem}
From Theorem~\ref{thm:PoA} it is clear that $\eta > 1.$ 

Let $U_{N}$ denote the sum utility of all the CPs in the {non-cooperative interaction}. Akin to the derivation of \eqref{eq:utility}, we obtain $U_{N}$ as
\begin{equation}
    U_{N} = \sum_{n \in \mathcal{N}} \left(r_{n}a_{n}\log(1+Q_{N}+b_{n}\sqrt{p_{n}}) - p_{n}\right) - Q_{N} . \label{eq:NE_sum_utility}
\end{equation}
Defining $\Gamma$ as 
\begin{equation}
    \Gamma := \frac{U_{C}^{\ast}}{U_{N}^{\ast}},  \label{eq:Gamma}
\end{equation}
we see from Theorem \ref{thm:PoA} that if $\max \limits_{n \in \mathcal{N}} \left(r_{n}a_{n} - \frac{b_{n}^2}{2}\right)>1$, $\Gamma>1$.
Thus, the CPs reduce their net utility in non-cooperative interaction by playing selfishly.
\section{Bargaining Game} \label{sec:Bargaining}
In this section, we consider that the CPs bargain with each other. In the case of disagreement, CPs do not make any public investment. Therefore, $Q=0$ is the point of disagreement. Let $U_{n}^{\ast,D}$ denote the optimum utility of the CP $n$ in case of disagreement and is obtained after optimizing (\ref{eq:Un}) over $p_{n}$ for $Q=0$ as
\begin{align}
U_{n}^{\ast,D} = r_{n}a_{n} \log\left(\frac{\sqrt{1+2b_{n}^{2}r_{n}a_{n}}\!+\!1}{2}\right)  - \left(\frac{\sqrt{1+2b_{n}^{2}r_{n}a_{n}}\!-\!1}{2b_{n}}\right)^{2} . \label{eq:Un_D}
\end{align}
Using (\ref{eq:Un_2}) and (\ref{eq:Un_D}), the set of feasible utility functions of the CPs, denoted by $\mathcal{U}$, is defined as
\begin{align}
    \mathcal{U} = \{(U_{1}(Q), \cdots, U_{N}(Q)) \,|\, Q \geq 0  \text{ and } U_{n}(Q) \geq U_{n}^{\ast,D} ~\forall n \in \mathcal{N}\}. \nonumber
\end{align}
A Nash bargaining solution $\textbf{U}^{B}=(U_{1}^{B}, \cdots, U_{N}^{B})$ is defined to be the solution of the problem formulated as
\begin{align}
    \max &\prod_{n \in \mathcal{N}} (U_{n} - U_{n}^{\ast, D}) \nonumber \\
    \text{s. t. } &(U_{1}, \cdots, U_{N}) \in \mathcal{U} . \label{eq:NBS}
\end{align}
The following lemma characterizes the condition for the existence of a non-zero solution of (\ref{eq:NBS}).
\begin{lemma}\label{lemma:NBS}
    A non-zero solution of Nash Bargaining problem given in (\ref{eq:NBS}) exists if and only if $\sum \limits_{n \in \mathcal{N}} \frac{\sqrt{1+2b_{n}^{2}r_{n}a_{n}}-1}{b_{n}^{2}} > 1$.
\end{lemma}
Given that a non-zero Nash bargaining solution exists, we present the following theorem that discusses its uniqueness. 
\begin{theorem}\label{thm:NBS}
    If $\sum_{n \in \mathcal{N}} \frac{\sqrt{1+2b_{n}^{2}r_{n}a_{n}}-1}{b_{n}^{2}} > 1$, then the solution of (\ref{eq:NBS}), denoted by $\textbf{U}^{B}=(U_{1}^{B}, U_{2}^{B}, \cdots, U_{N}^{B})$, is unique.
\end{theorem}
Let $Q_{B}^{\ast}$ denote the optimum total public investment in the bargaining game. From Lemma \ref{lemma:NBS} and Theorem \ref{thm:NBS}, $Q_{B}^{\ast}$ is unique and non-zero if and only if $\sum_{n \in \mathcal{N}} \frac{\sqrt{1+2b_{n}^{2}r_{n}a_{n}}-1}{b_{n}^{2}} > 1$. This condition is the same as that required for a positive value of optimum total public investment in the centralized allocation, i.e., $Q_{C}^{\ast}$ (Refer to Lemma \ref{lem:lemma_coop}). Thus, we present the following theorem that explores the relation between the optimum total public investment in the bargaining game and centralized allocation.  
\begin{theorem}\label{thm:NBS2}
\color{white} dummy text \color{black}
\begin{itemize} 
    \item If $\sum_{n \in \mathcal{N}} \frac{\sqrt{1+2b_{n}^{2}r_{n}a_{n}}-1}{b_{n}^{2}}>1$, then a unique $Q_{B}^{\ast}$ and $q_{1}^{B}, \cdots, q_{N}^{B} \geq 0$ exist such that 
    \begin{equation}
        \sum_{n=1}^{N} q_{n}^{B} = Q_{B}^{\ast} \text{ and } U_{n}^{B} = U_{n}(Q_{B}^{\ast}) , \forall n \in \mathcal{N}. \nonumber
    \end{equation}
    Further, $Q_{B}^{\ast}>Q_{C}^{\ast}$.
    \item If $\sum_{n \in \mathcal{N}} \frac{\sqrt{1+2b_{n}^{2}r_{n}a_{n}}-1}{b_{n}^{2}}>1$ and $q_{n}^{B}>0$ for all $n \in \mathcal{N}$, then $Q_{B}^{\ast}=Q_{C}^{\ast}$.
\end{itemize}
\end{theorem}
Let $\beta$ denote the benefit of bargaining which is defined as
\begin{equation}
    \beta := \frac{Q_{B}^{\ast}}{Q_{C}^{\ast}} . \label{eq:beta}
\end{equation}
From Theorem \ref{thm:NBS2}, $\beta \geq 1$ under certain conditions. This implies that the bargaining game can result in higher public investment than the centralized allocation. 
However, note that the net utility of the CPs in the bargaining game can not be more than that in the centralized allocation.
This follows from the concavity of the utility function.
Further, note that all the CPs invest in their private infrastructure, and the optimum private investment of CP $n$, i.e., $p_{n}^{\ast}$, can be obtained by substituting $Q$ by $Q_{B}^{\ast}$ in (\ref{eq:opti_pn}). Then, $\gamma_{B}$ denotes the public-private trade-off, defined in (\ref{eq:gamma}), for the bargaining game. 

We use $\alpha$ to denote the product of the price of anarchy obtained in the non-cooperative game and the benefit of bargaining obtained in this game. Thus, $\alpha=\eta \times \beta$. From (\ref{eq:eta}) and (\ref{eq:beta}), note that $\alpha$ compares the optimum total public investment in the non-cooperative game and this game. We will evaluate the impact of system parameters on $\gamma_{B}$ and $\alpha$ in Section \ref{sec:Numericals}.
\section{Numerical Results}
\label{sec:Numericals}
\begin{figure*}
    \centering
    \subfigure[]
    {
        \includegraphics[height=2in,width=3in]{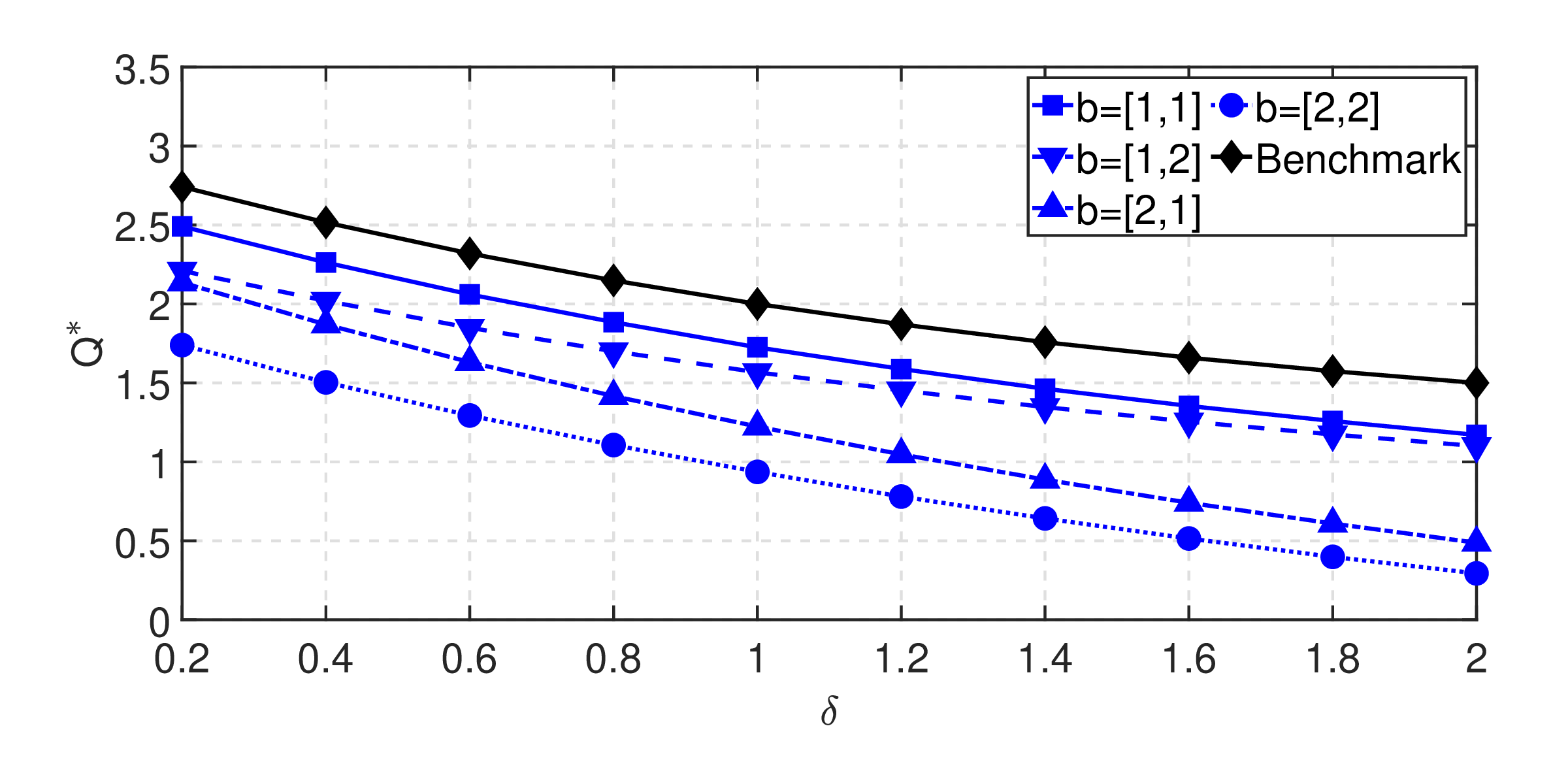}
        \label{fig:cent_pub}
    }
    \subfigure[]
    {
        \includegraphics[height=2in,width=3in]{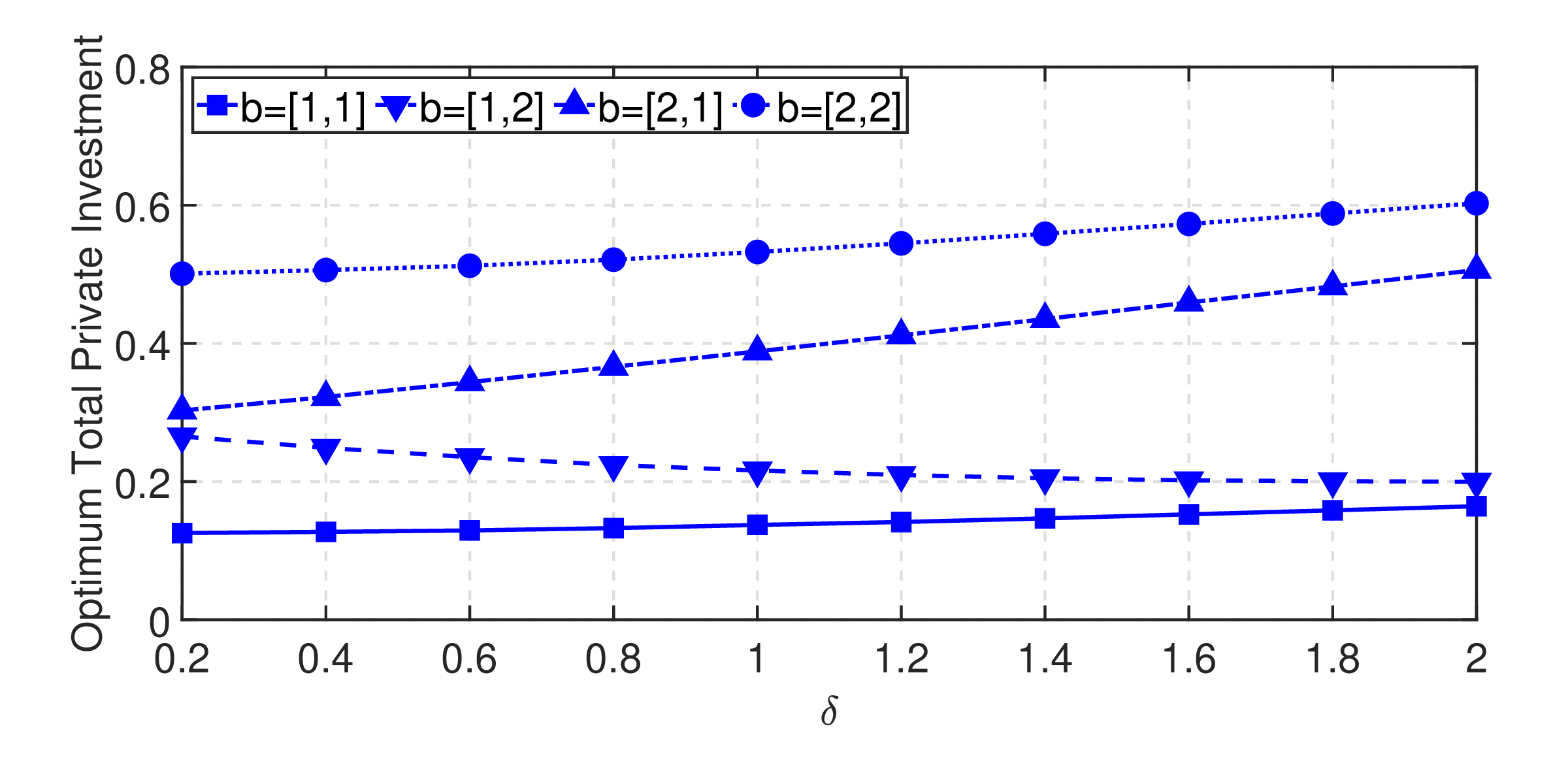}
        \label{fig:cent_priv}
    }
    \subfigure[]
    {
        \includegraphics[height=2in,width=3in]{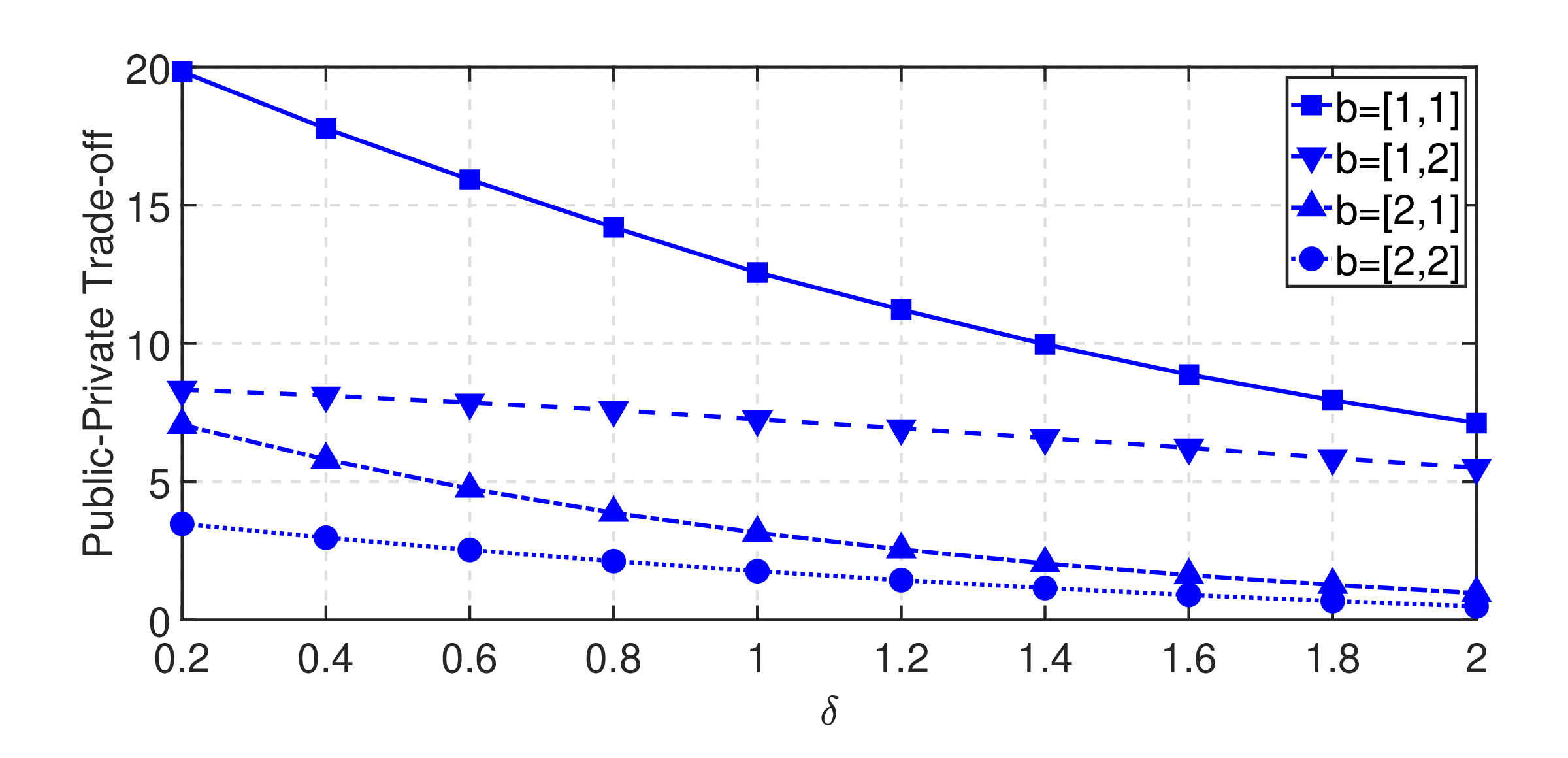}
        \label{fig:cent_ppt}
    }
    \subfigure[]
    {
        \includegraphics[height=2in,width=3in]{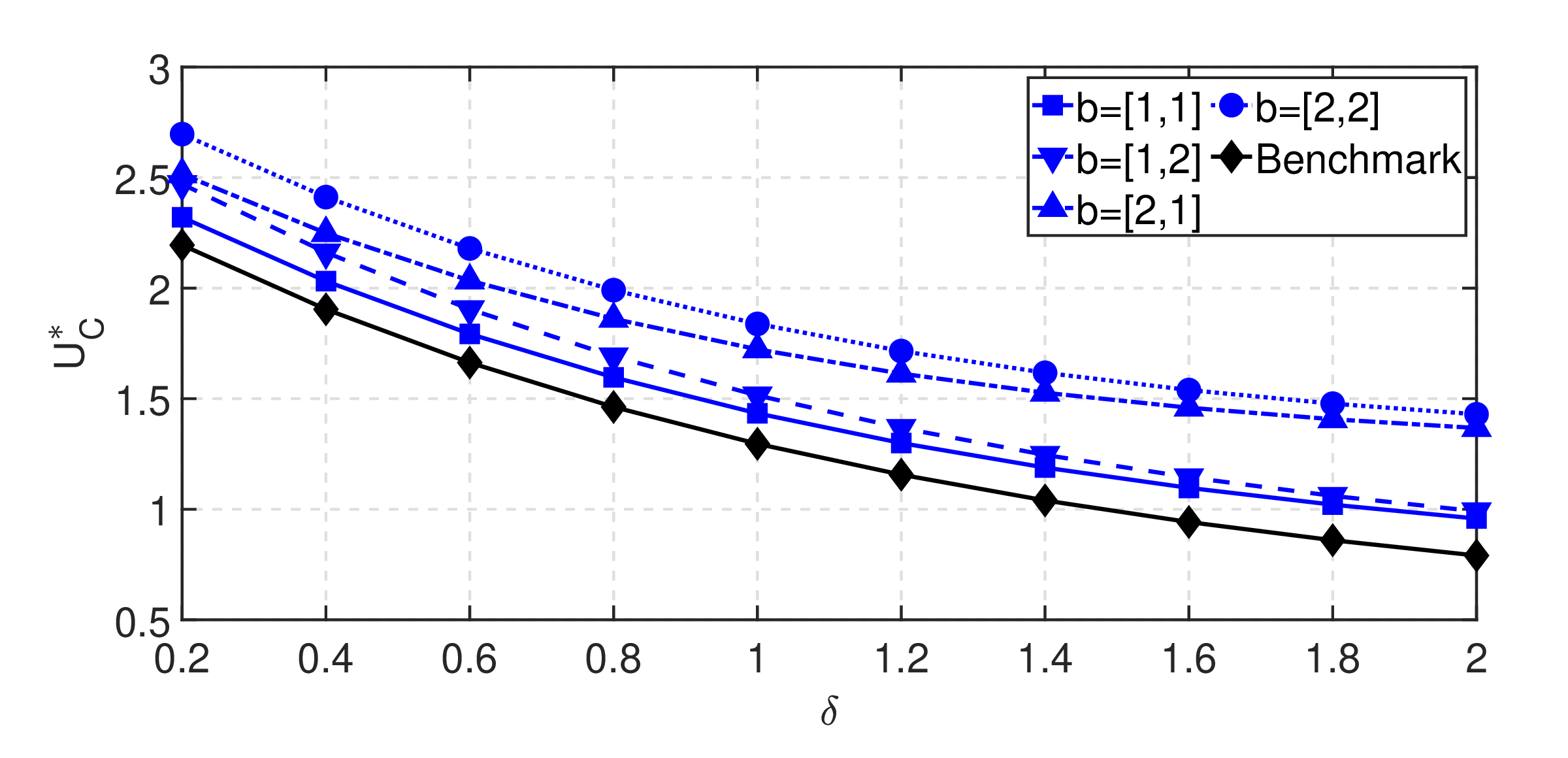}
        \label{fig:cent_util}
    }
    \caption{Variation of optimum total public investment, optimum total private investment, public-private trade-off, and optimum total utility against different values of $\delta$ and $\textbf{b}$ for centralized allocation and $N=2$ in (a), (b), (c), and (d), respectively.
    }
    \label{fig:centralized}
\end{figure*}
In this section, we present some numerical results for $N=2.$ We consider $\psi_{n}=r_{n}a_{n} = 2n^{-\delta}$ for all $n=\{1,2\}.$ Here, $\delta$ controls the revenue of a CP per unit of traffic, and the multiplicative factor of $2$ ensures that the optimum total public investment is non-zero for all values of $\delta$, i.e., \eqref{eq:non-zero_cond} holds for all $\delta$. Note that the second CP's revenue per traffic unit diminishes as $\delta$ increases. 

Fig. \ref{fig:centralized} presents the variation of the optimum total public investment, optimum total private investment, public-private trade-off, and optimum total utility against different values of $\delta$ and $\textbf{b}$ for {centralized} allocation in Fig. \ref{fig:cent_pub}, \ref{fig:cent_priv}, \ref{fig:cent_ppt}, and \ref{fig:cent_util}, respectively. 
Note that the `Benchmark' in Fig. \ref{fig:cent_pub} and \ref{fig:cent_util} refer to the scenario considered in \citet{Kalvit2019}, wherein the CPs do not have an option of private investment and thus can only make public investment.
We observe from Fig. \ref{fig:cent_pub} that the optimum total public investment, i.e., $Q_{C}^{\ast}$, decreases as $\delta$ increases. This follows from the fact that the revenue of CP 2 per unit of traffic is a decreasing function of $\delta$. Since the revenue of CP 1 per unit of traffic does not change with $\delta$, the decrease in $Q_{C}^{\ast}$ leads to an increase in the optimum private investment of CP 1 (Refer Lemma \ref{lem:opti_pn}). However, since the revenue of CP 2 per unit of traffic decreases with $\delta$, the optimum private investment of CP 2 decreases with $\delta$. This explains the mixed trend of the optimum total private investment, which is the sum of the optimal private investments of both CPs, against different values of $\delta$ as observed in Fig. \ref{fig:cent_priv}. In the case of $\textbf{b}=[2, 2]$, both the CPs are incentivized to invest more in private than public infrastructure. Therefore, we observe that the optimal total public and private investment are lowest and highest for $\textbf{b}=[2, 2]$ in Fig. \ref{fig:cent_pub} and \ref{fig:cent_priv}, respectively. A similar explanation holds for the highest and lowest values of optimum total public and private investments for the case of $\textbf{b}=[1, 1]$ in Fig. \ref{fig:cent_pub} and \ref{fig:cent_priv}, respectively. 
This also translates into the highest public-private trade-off for $\textbf{b}=[1, 1]$ in Fig. \ref{fig:cent_ppt}.
The higher private investment in the case of $\textbf{b}=[2, 2]$ provides an additional boost to the individual utilities of the CPs. Thus, we observe from Fig. \ref{fig:cent_util} that the net utility is the highest for the case of $\textbf{b}=[2, 2]$.
CP 1 earns more revenue per unit of traffic than CP 2 for all values of $\delta$. Further, unlike CP 2, the investment of CP 1 is not skewed towards the private infrastructure in the case of $\textbf{b}=[1, 2]$. Thus, we observe higher value of $Q_{C}^{\ast}$ for $\textbf{b}=[1, 2]$ than $\textbf{b}=[2, 1]$ in Fig. \ref{fig:cent_pub}.
The observation of lower optimum total private investment for $\textbf{b}=[1, 2]$ than $\textbf{b}=[2,1]$ in Fig. \ref{fig:cent_priv} also follows from the similar explanation. 
However, an interesting observation from Fig. \ref{fig:cent_util} is that the net utility is higher for $\textbf{b}=[2, 1]$ than $\textbf{b}=[1, 2]$. This follows from two facts. First, CP 1 dominates the net utility as it has higher revenue per unit of traffic than CP 2. Secondly, the incentive to invest more in private than public for CP 1 provides an additional boost to its utility and, in turn, the net utility.
For all the cases except $\textbf{b}=[1, 1]$, CP(s) can obtain higher benefits from private investment, leading to the reduction in the public-private trade-off, as observed from Fig. \ref{fig:cent_ppt}.
We observe from Fig. \ref{fig:cent_pub} that the optimum total public investment for the benchmark model considered in \citet{Kalvit2019} is higher than that obtained in the proposed model for all values of $\textbf{b}$. However, despite that, the proposed model obtains higher net utility than the benchmark model for all values of $\textbf{b}$ (Refer Fig. \ref{fig:cent_util}). This results from the additional gain obtained from the private investment. 
\begin{figure*}
    \centering
    \subfigure[]
    {
        \includegraphics[height=1.6in,width=2in]{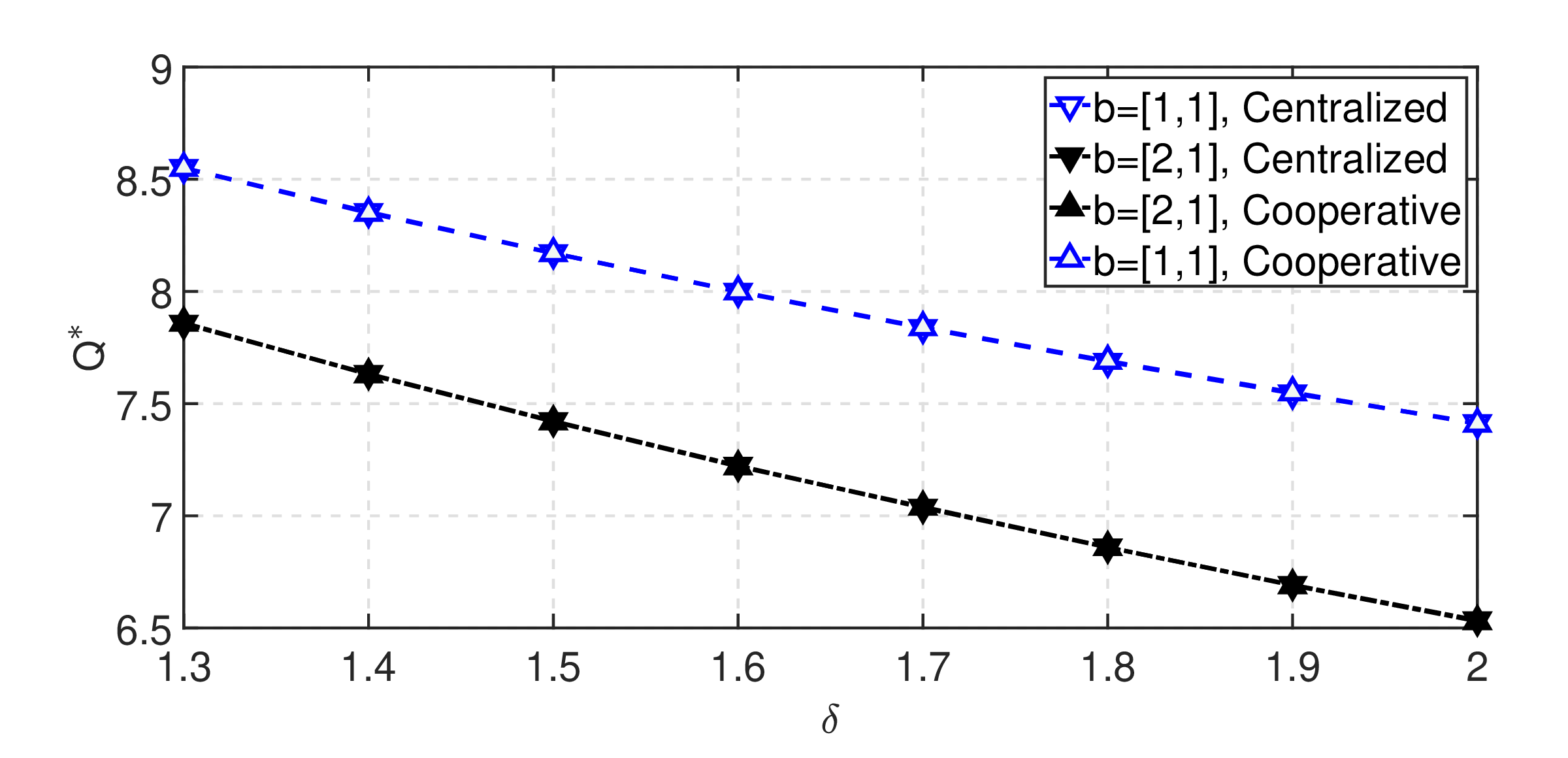}
        \label{fig:coop_pub}
    }
    \subfigure[]
    {
        \includegraphics[height=1.6in,width=2in]{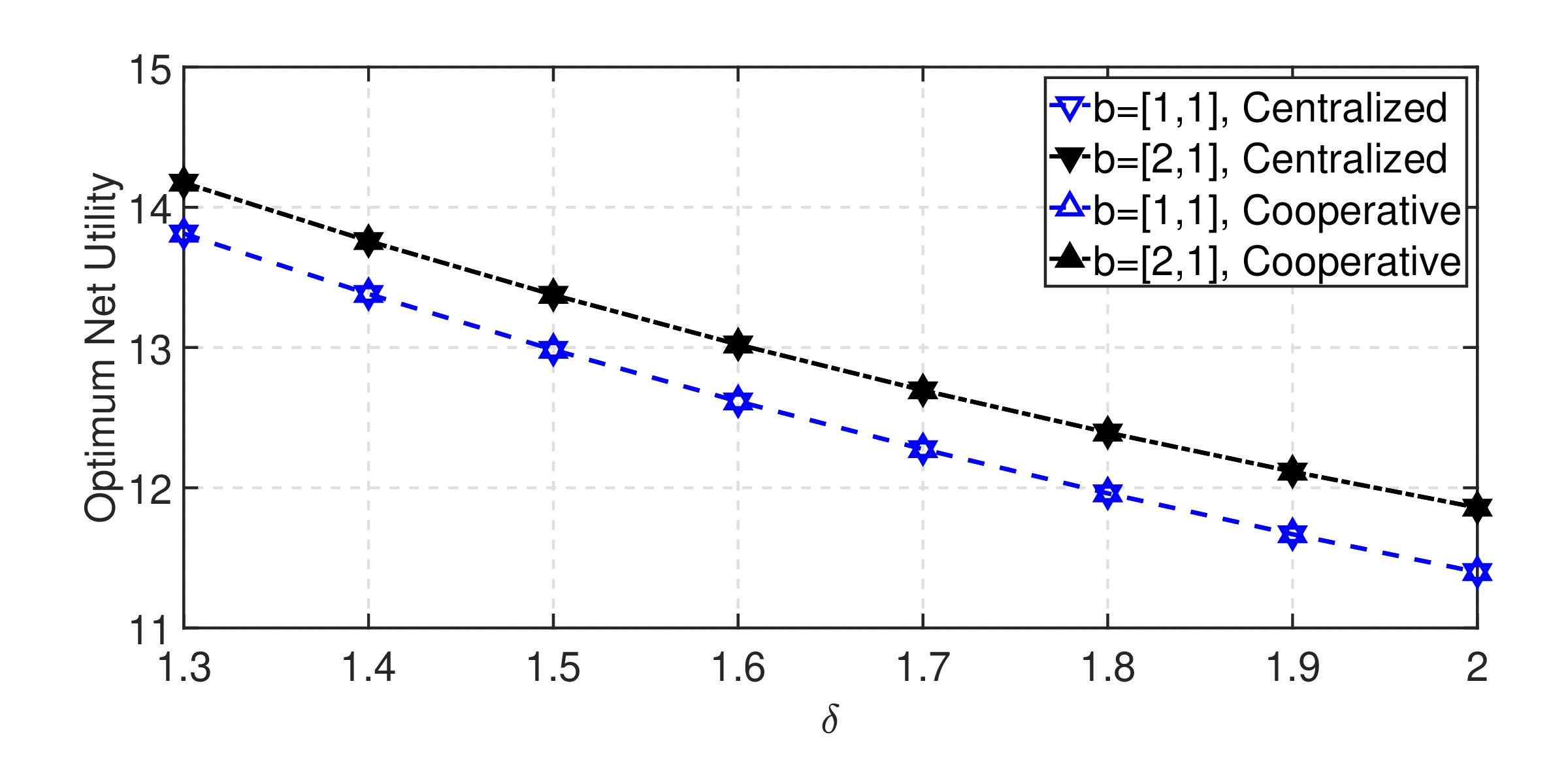}
        \label{fig:coop_util}
    }
    \subfigure[]
    {
        \includegraphics[height=1.6in,width=2in]{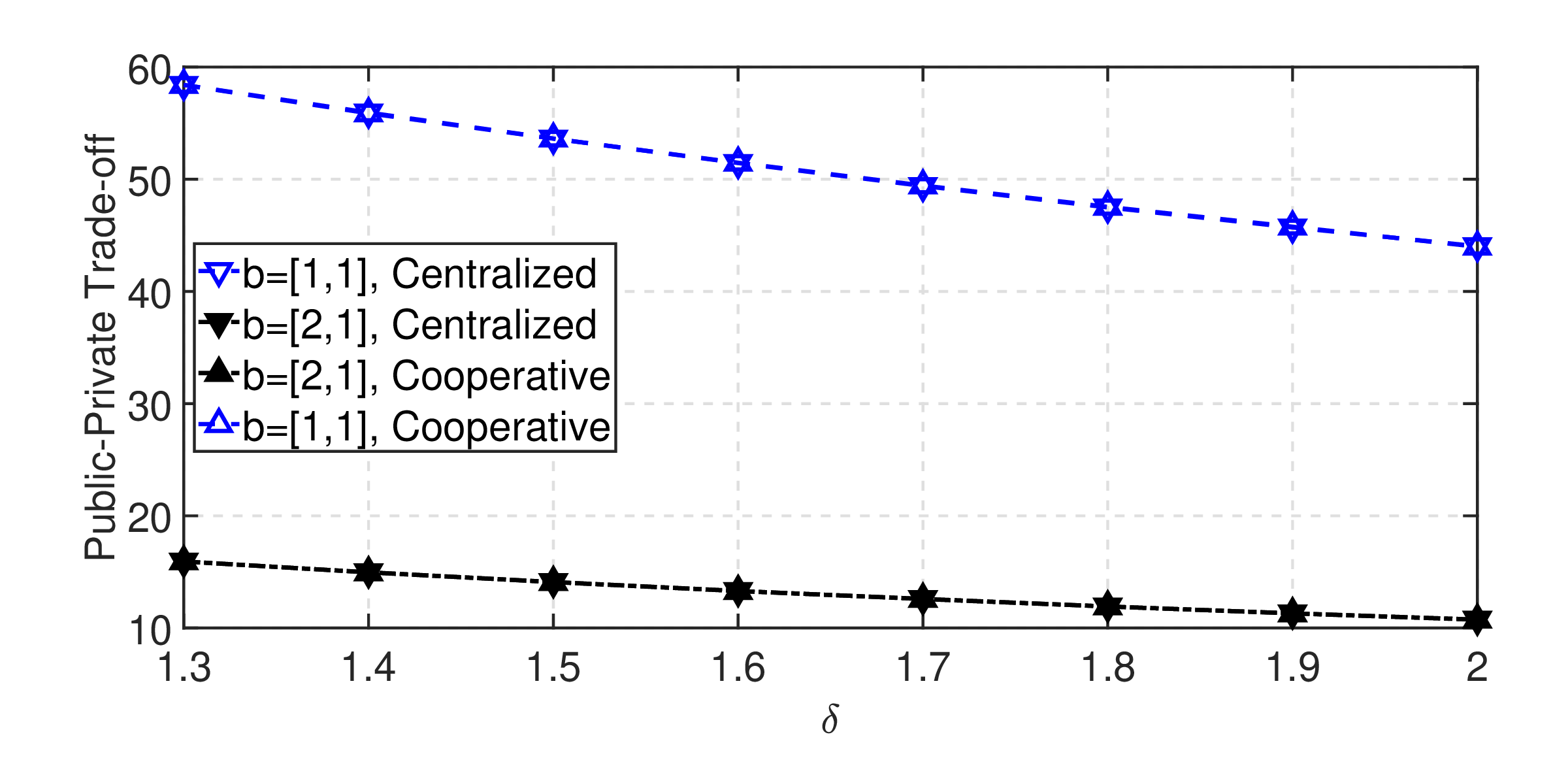}
        \label{fig:coop_ppt}
    }
    \caption{Variation of optimum total public investment, optimum net utility, and public-private trade-off against different values of $\delta$ and $\textbf{b}$ for centralized allocation and cooperative game and $N=2$ in (a), (b), and (c), respectively.
    }
    \label{fig:cooperative}
\end{figure*}
\begin{figure*}
    \centering
    \subfigure[]
    {
        \includegraphics[height=1.6in,width=2in]{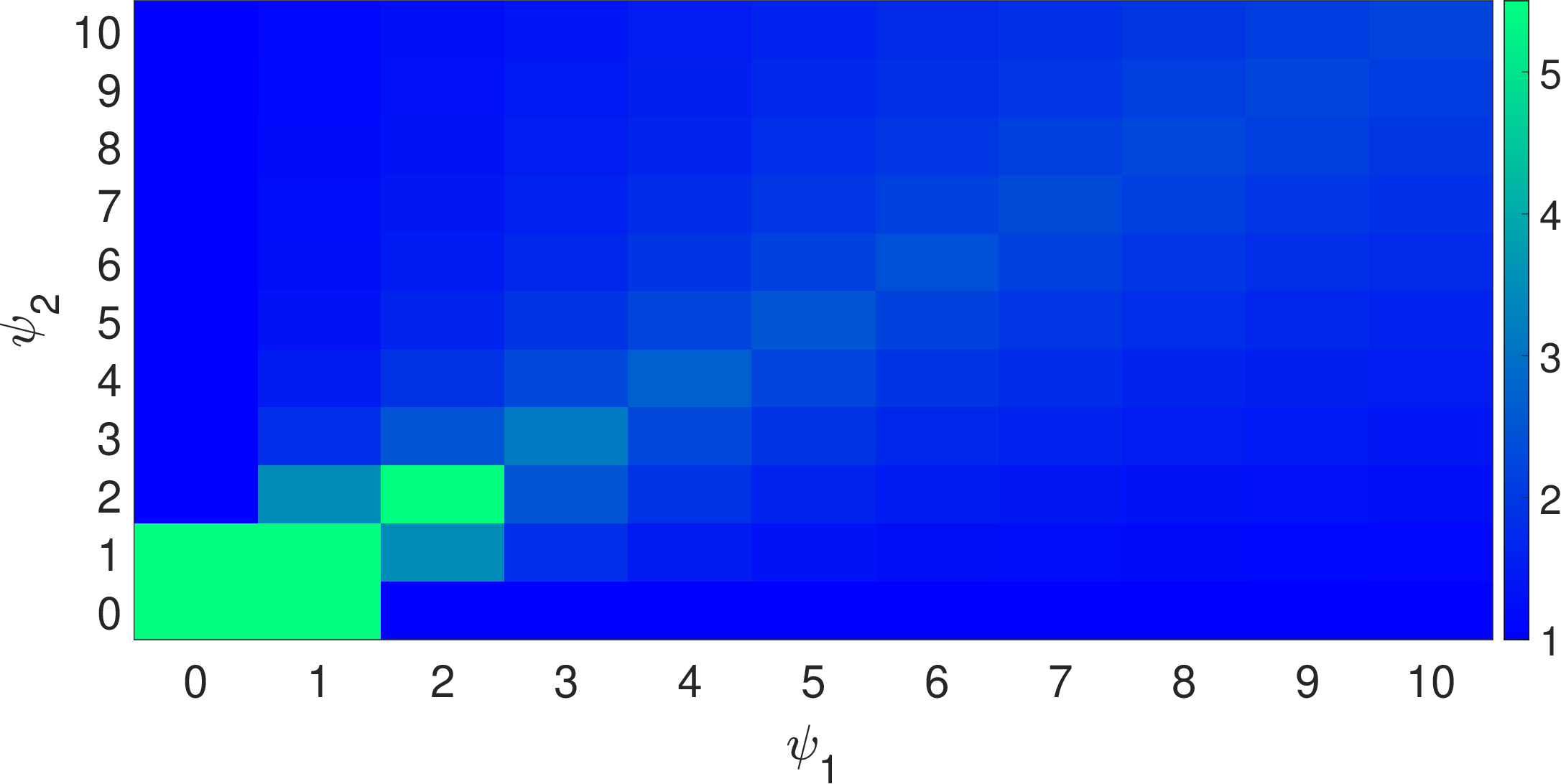}
        \label{fig:poa}
    }
    \subfigure[]
    {
        \includegraphics[height=1.6in,width=2in]{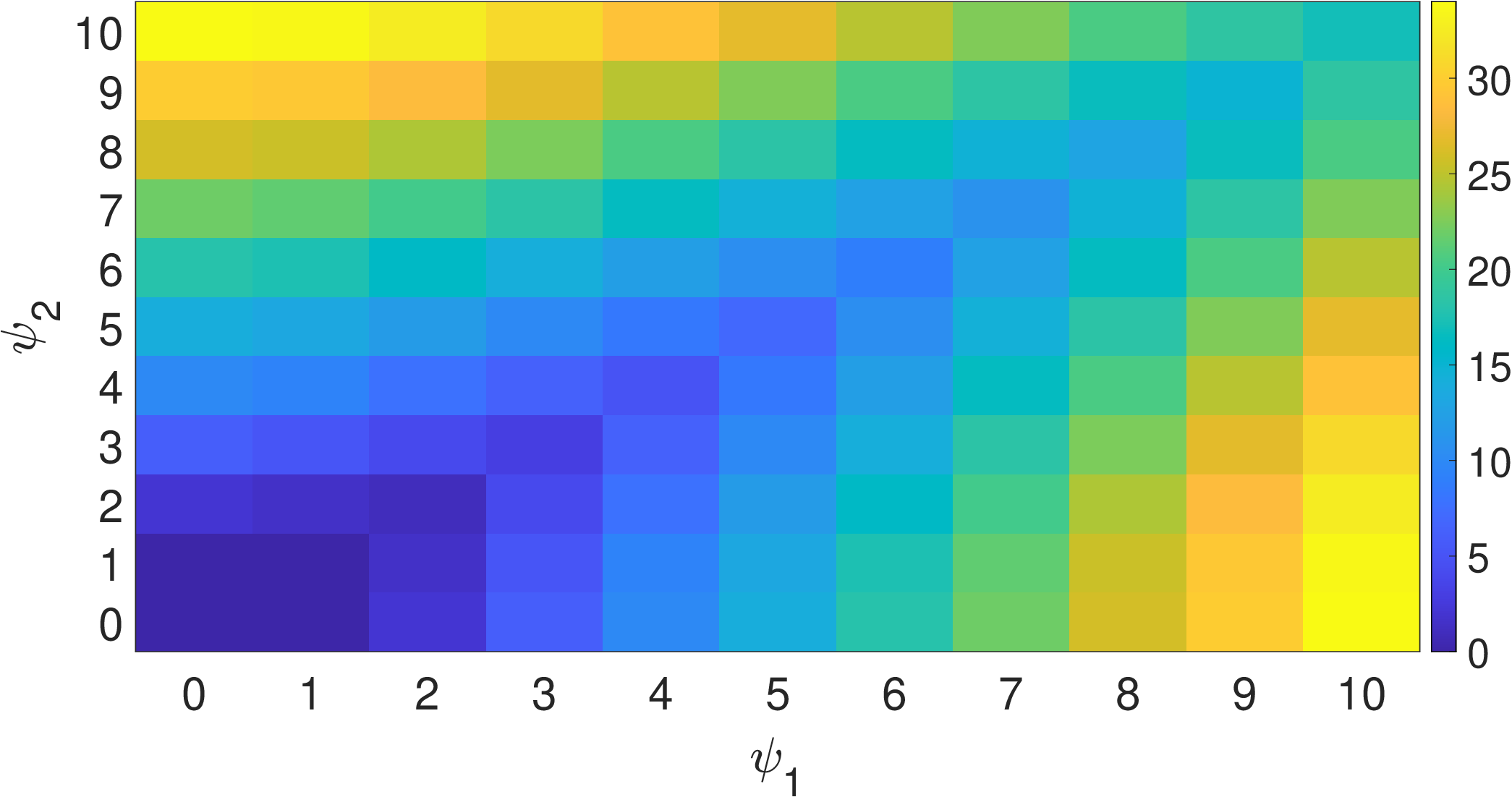}
        \label{fig:NE_ppt}
    }
    \subfigure[]
    {
        \includegraphics[height=1.6in,width=2in]{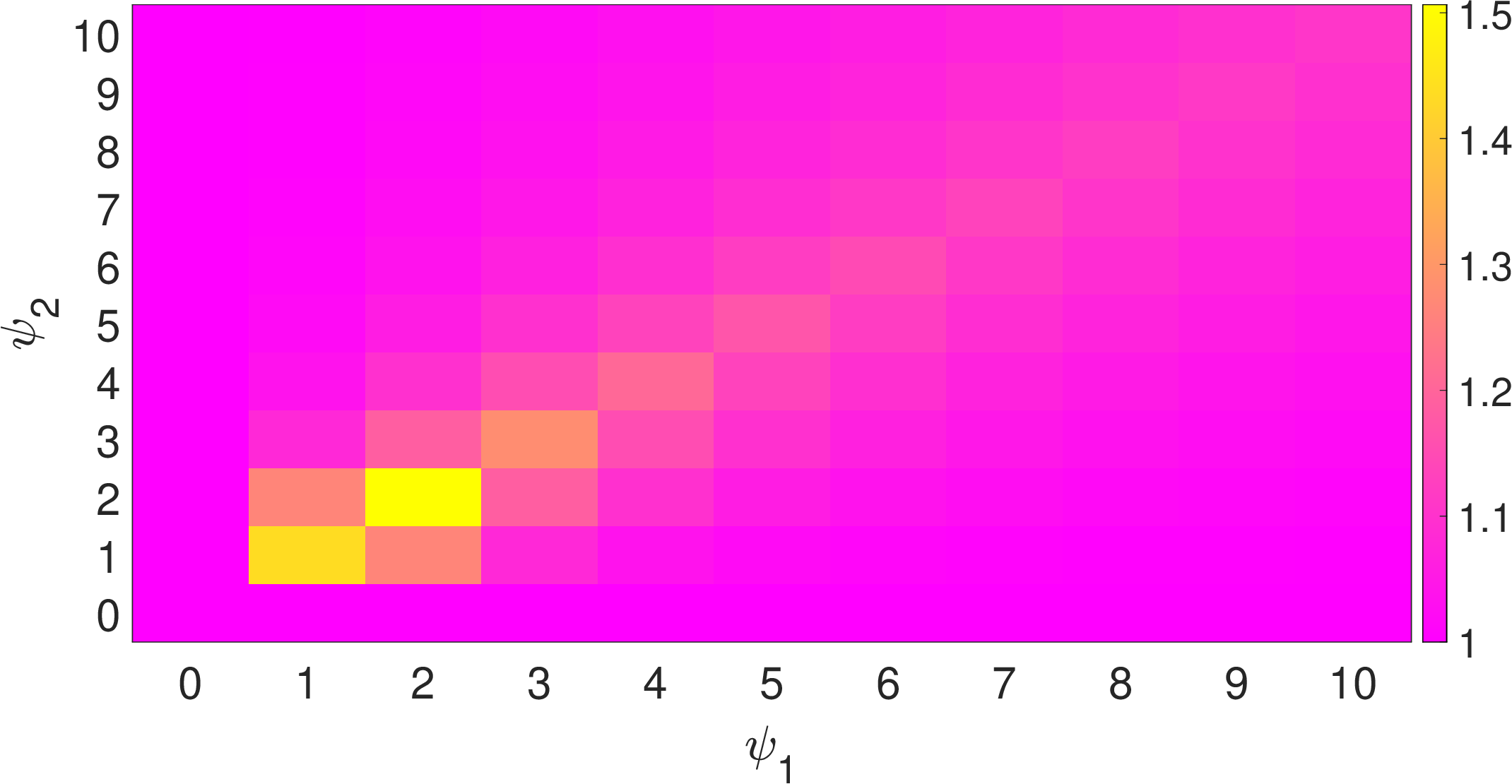}
        \label{fig:gamma}
    }
    \caption{Illustration of the price of anarchy ($\eta$), the public-private trade-off ($\gamma_{N}$), and $\Gamma$ against different values of $\psi_{n}(=r_{n}a_{n})$ for non-cooperative game for $N=2$ and $\textbf{b}=[1, 1]$ in (a), (b), and (c), respectively.
    }
    \label{fig:NE}
\end{figure*}

Fig. \ref{fig:cooperative} presents the variation of optimum total public investment, optimum net utility, and public-private trade-off against different values of $\delta$ and $\textbf{b}$ for centralized allocation and cooperative game in Fig. \ref{fig:coop_pub}, \ref{fig:coop_util}, and \ref{fig:coop_ppt}, respectively. For these numerical results, we consider $\psi_{n}=r_{n}a_{n}=7n^{-\delta}$ for all $n \in \{1,2\}$. Further, the numerical results for centralized allocation and cooperative game are obtained by solving (\ref{eq:centralized_problem}) and (\ref{eq:coalition_problem}), respectively. We observe from Fig. \ref{fig:cooperative} that the optimum total public investment and optimum net utility obtained in the cooperative game are the same as those obtained from centralized allocation. The optimum total public investment is less in $\textbf{b}=[2,1]$ than $\textbf{b}=[1,1]$ as CP 1 has an incentive to invest more in private. This also benefits the utility of CP 1, and hence, we observe higher optimum net utility in $\textbf{b}=[2,1]$ than $\textbf{b}=[1,1]$ in Fig. \ref{fig:coop_util}. The reduction in the public-private trade-off in $\textbf{b}=[2,1]$ than $\textbf{b}[1,1]$, as observed in Fig. \ref{fig:coop_ppt}, also follows from the above discussion. A key observation from Fig. \ref{fig:cooperative} is that the grand coalition of all the CPs is stable for certain scenarios even when CP(s) can free-ride on the investment of other CPs.  

\begin{figure*}
    \centering
    \subfigure[]
    {
        \includegraphics[height=1.6in,width=2in]{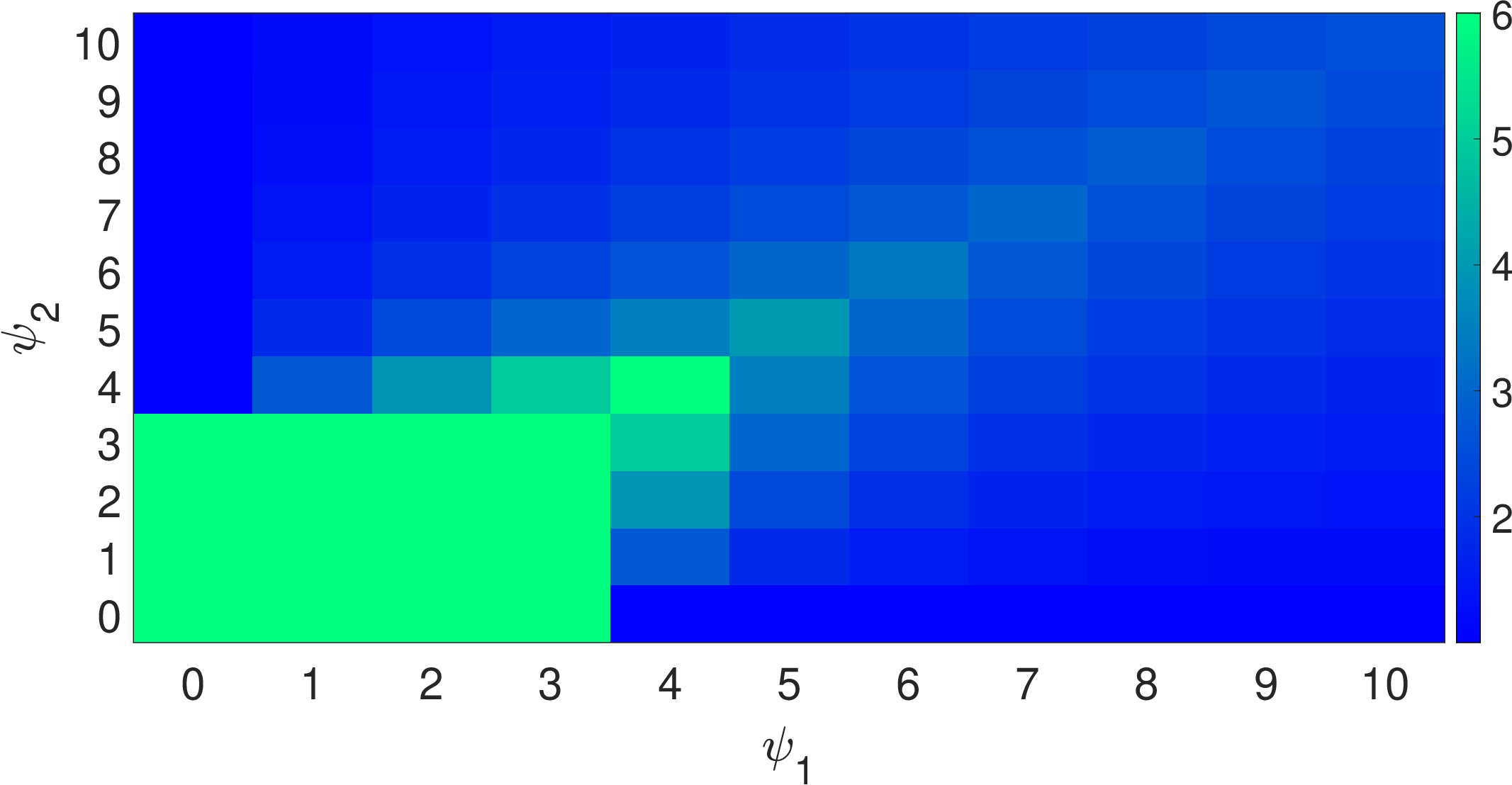}
        \label{fig:poa_b22}
    }
    \subfigure[]
    {
        \includegraphics[height=1.6in,width=2in]{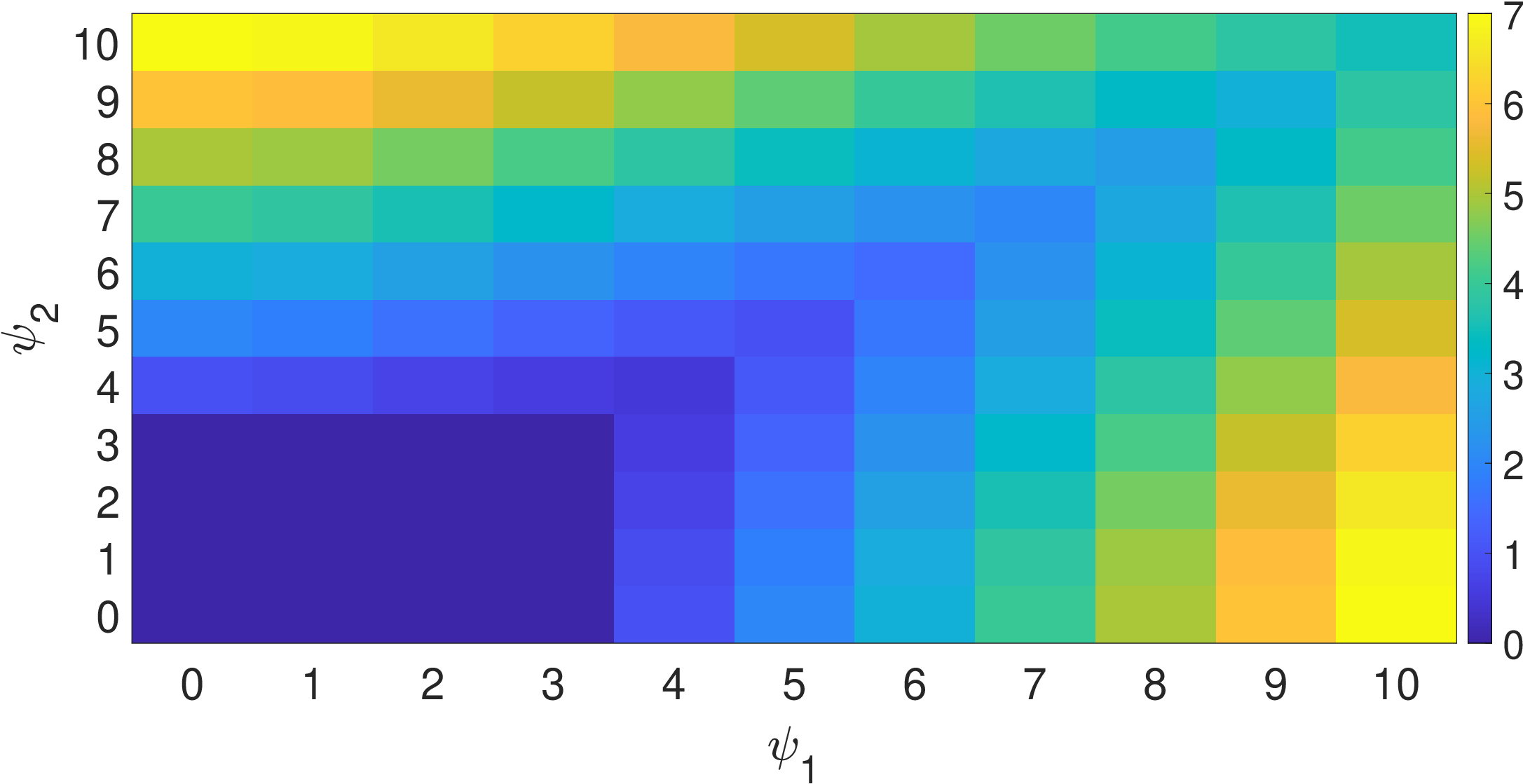}
        \label{fig:NE_ppt_b22}
    }
    \subfigure[]
    {
        \includegraphics[height=1.6in,width=2in]{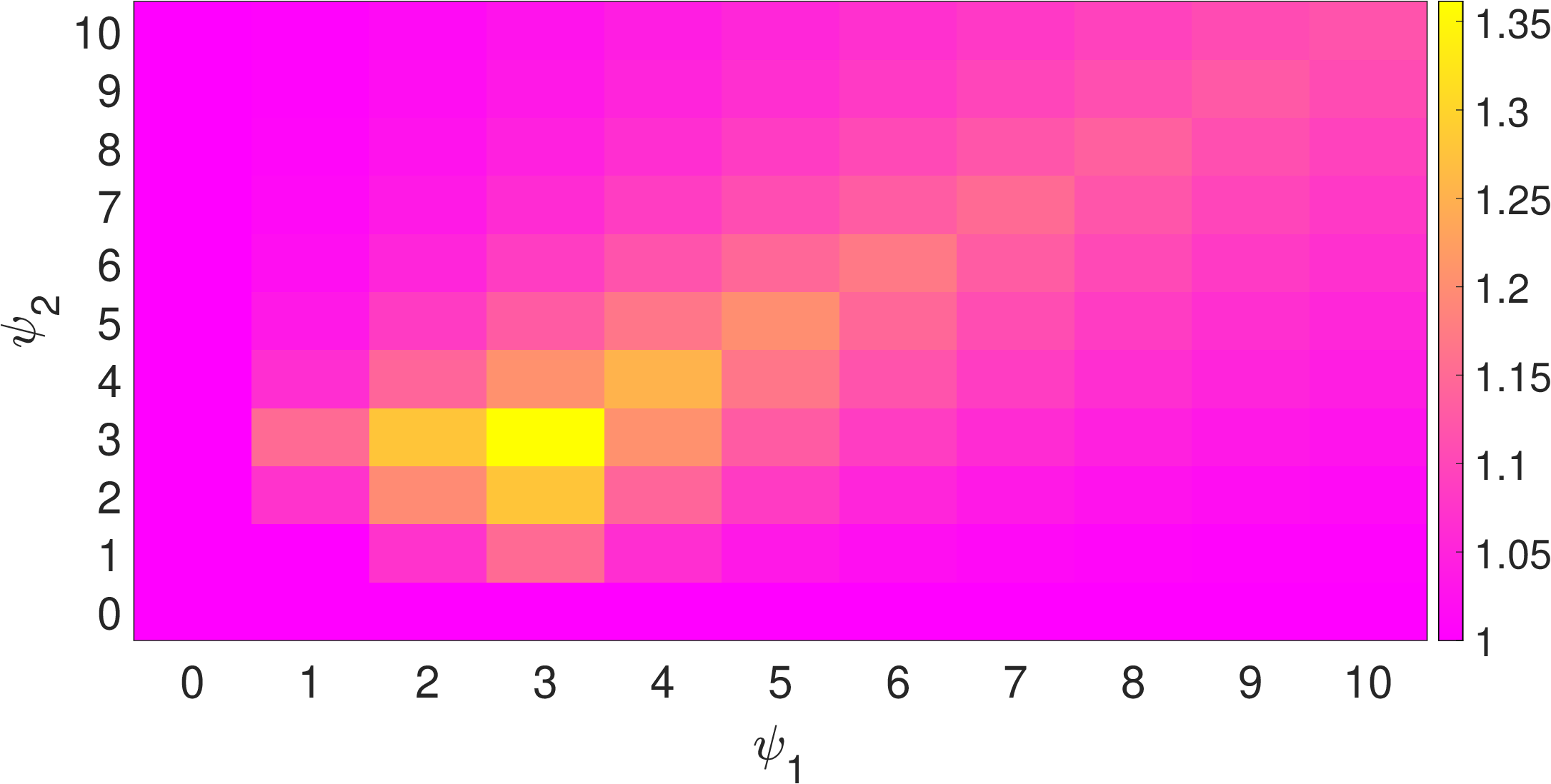}
        \label{fig:gamma_b22}
    }
    \caption{Illustration of the price of anarchy ($\eta$), the public-private trade-off ($\gamma_{N}$), and $\Gamma$ against different values of $\psi_{n}(=r_{n}a_{n})$ for non-cooperative game for $N=2$ and $\textbf{b}=[2, 2]$ in (a), (b), and (c), respectively.
    }
    \label{fig:NE_b22}
\end{figure*}

Fig. \ref{fig:NE} illustrates price of anarchy ($\eta$), the public-private trade-off ($\gamma_{N}$), and $\Gamma$ (defined in (\ref{eq:Gamma})) against different values of $\psi_{n}(=r_{n}a_{n})$ for non-cooperative game for $\textbf{b}=[1, 1]$ in Fig. \ref{fig:poa}, \ref{fig:NE_ppt}, and \ref{fig:gamma}, respectively. 
Similarly, Fig. \ref{fig:NE_b22} illustrates price of anarchy ($\eta$), the public-private trade-off ($\gamma_{N}$) , and $\Gamma$ against different values of $\psi_{n}(=r_{n}a_{n})$ for non-cooperative game for $\textbf{b}=[2, 2]$ in Fig. \ref{fig:poa_b22}, \ref{fig:NE_ppt_b22}, and \ref{fig:gamma_b22}, respectively. 
We observe from Fig. \ref{fig:poa} that $\eta$ is high for the region $(\psi_{1} \leq 1.5, \psi_{2} \leq 1.5)$. This follows from Theorem \ref{thm:PoA} as $\max \limits_{n \in \mathcal{N}} \left(r_{n}a_{n} - \frac{b_{n}^{2}}{2}\right) \leq 1$ in this region. 
Same holds for the region $(\psi_{1} \leq 3, \psi_{2} \leq 3)$ in Fig. \ref{fig:poa_b22} as $\textbf{b}=[2, 2]$. 
In the remaining region, we observe $\eta>1$. It is consistent with Theorem \ref{thm:PoA}. When $\psi_{1} \approx \psi_{2}$, both CPs invest in public infrastructure leading to a higher value of $Q_{C}^{\ast}$, and hence, $\eta$. However, when there is a large asymmetry between the CPs, only the dominating CP makes a major investment in public infrastructure, leading to a lower value of $\eta$.
Similar observations can be made from Fig. \ref{fig:poa_b22}. However, note that the optimum total public investment is less in the case of $\textbf{b}=[2,2]$ than $\textbf{b}=[1,1]$ for both centralized allocation and non-cooperative game. This is because both the CPs are incentivized to invest more in private than public. Thus, we observe that $\eta$ is slightly higher for $\textbf{b}=[2, 2]$ than $\textbf{b}=[1, 1]$.
The same reason holds for lower value of $\gamma_{N}$ for $\textbf{b}=[2,2]$ than $\textbf{b}=[1,1]$, as observed from Fig. \ref{fig:NE_ppt} and \ref{fig:NE_ppt_b22}. Further, as the revenue per unit of traffic increases, $Q_{N}^{\ast}$ increases. Thus, the optimum total private investment decreases, and $\gamma_{N}$ increases as observed in Fig. \ref{fig:NE_ppt} and \ref{fig:NE_ppt_b22}.
In centralized allocation, the optimum total public investment maximizes the net utility of all the CPs. However, in the non-cooperative game, CPs play selfishly and reduce the net utility as observed in Fig. \ref{fig:gamma}. Same holds for the case of $\textbf{b}=[2, 2]$ as observed in Fig. \ref{fig:gamma_b22}.
\begin{figure*}
    \centering
    \subfigure[]
    {
        \includegraphics[height=1.6in,width=2in]{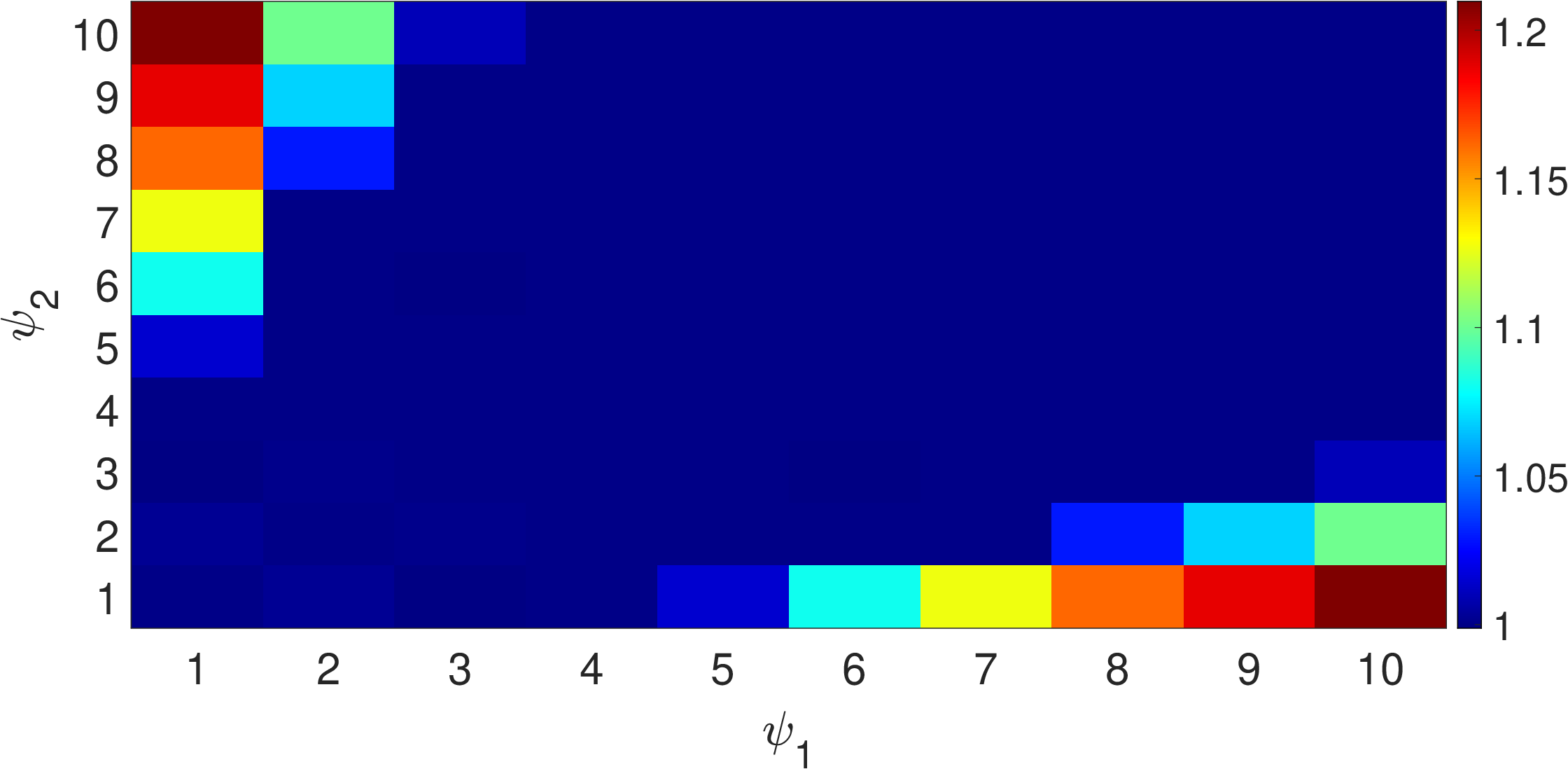}
        \label{fig:bob_b11}
    }
    \subfigure[]
    {
        \includegraphics[height=1.6in,width=2in]{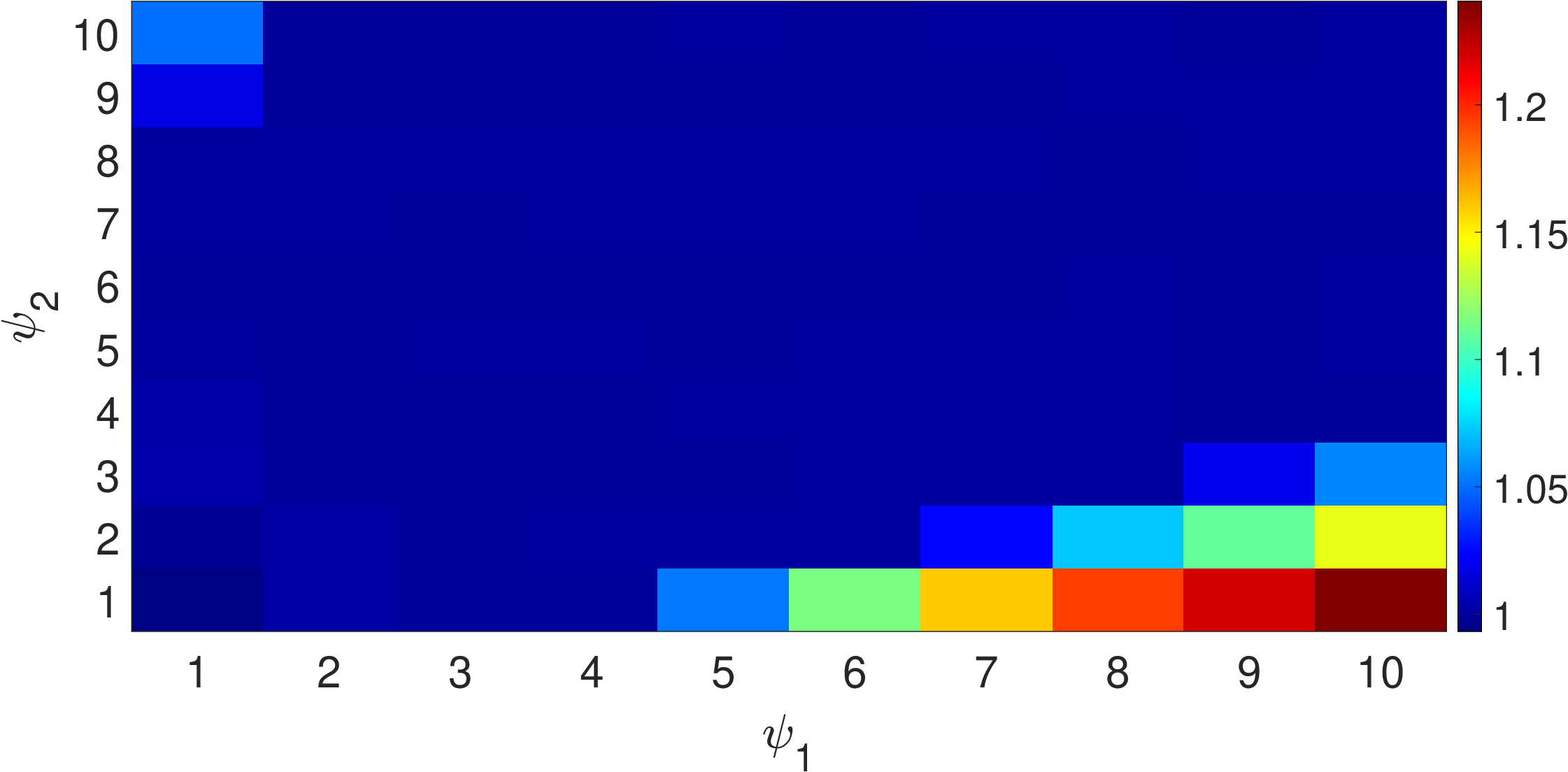}
        \label{fig:bob_b12}
    }
    \subfigure[]
    {
        \includegraphics[height=1.6in,width=2in]{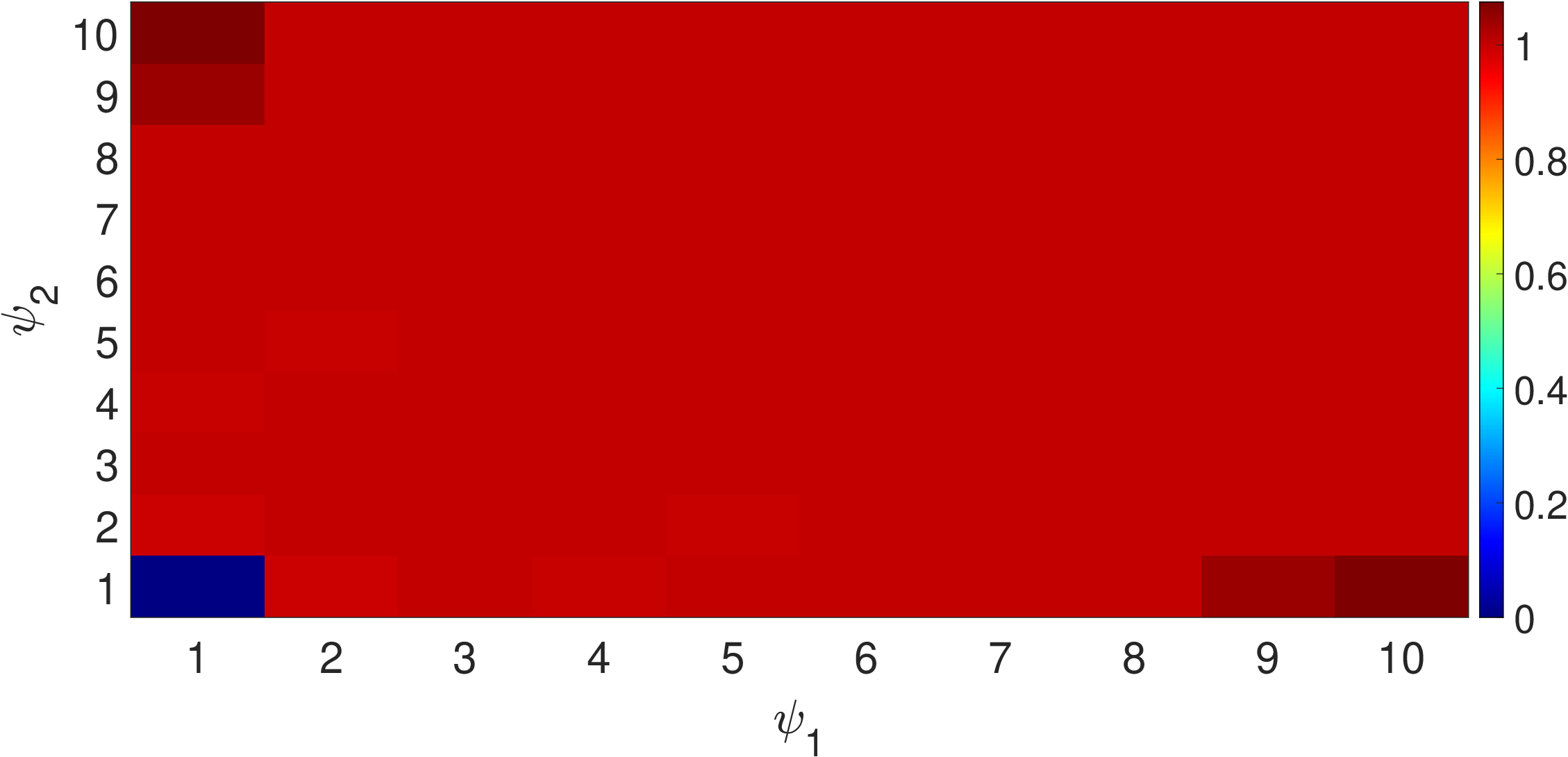}
        \label{fig:bob_b22}
    }
    \caption{Illustration of the benefit of bargaining ($\beta$) for bargaining game against different values of $\psi_{n}(=r_{n}a_{n})$ for $N=2$ for $\textbf{b}=[1, 1]$, $\textbf{b}=[1, 2]$, and $\textbf{b}=[2, 2]$ in (a), (b), and (c), respectively.
    }
    \label{fig:NB_BoB}
\end{figure*}
\begin{figure*}
    \centering
    \subfigure[]
    {
        \includegraphics[height=1.6in,width=2in]{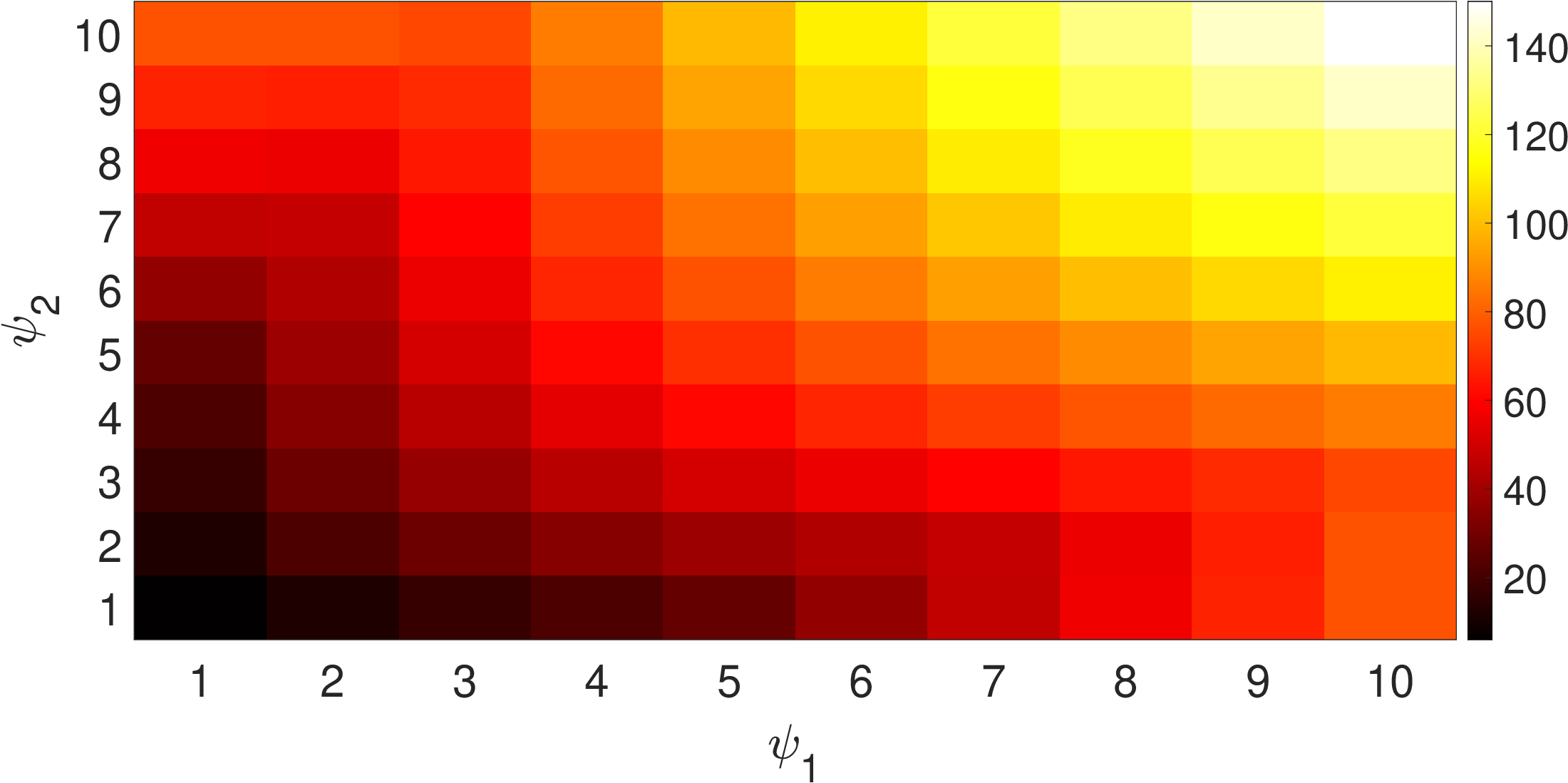}
        \label{fig:nb_ppt_b11}
    }
    \subfigure[]
    {
        \includegraphics[height=1.6in,width=2in]{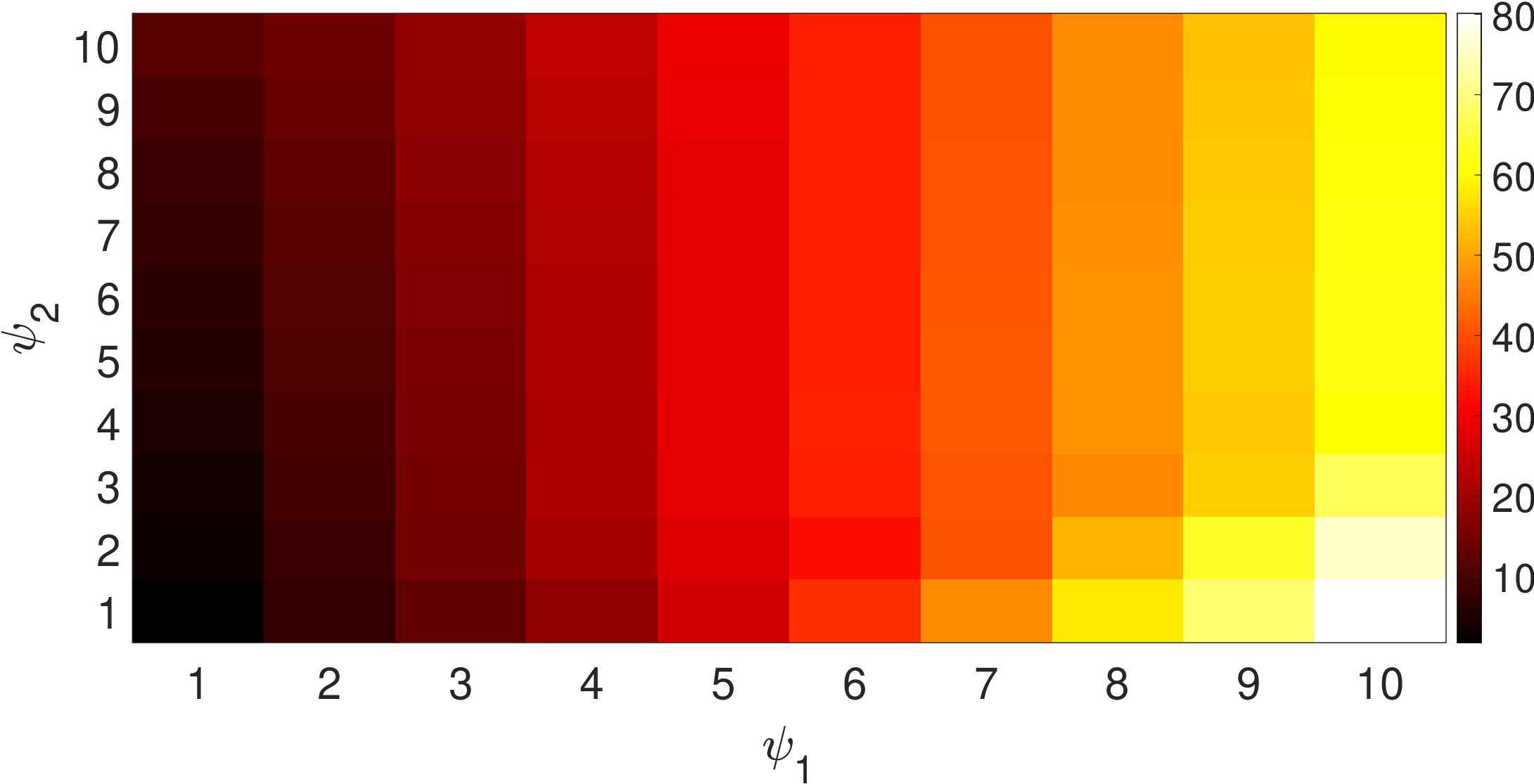}
        \label{fig:nb_ppt_b12}
    }
    \subfigure[]
    {
        \includegraphics[height=1.6in,width=2in]{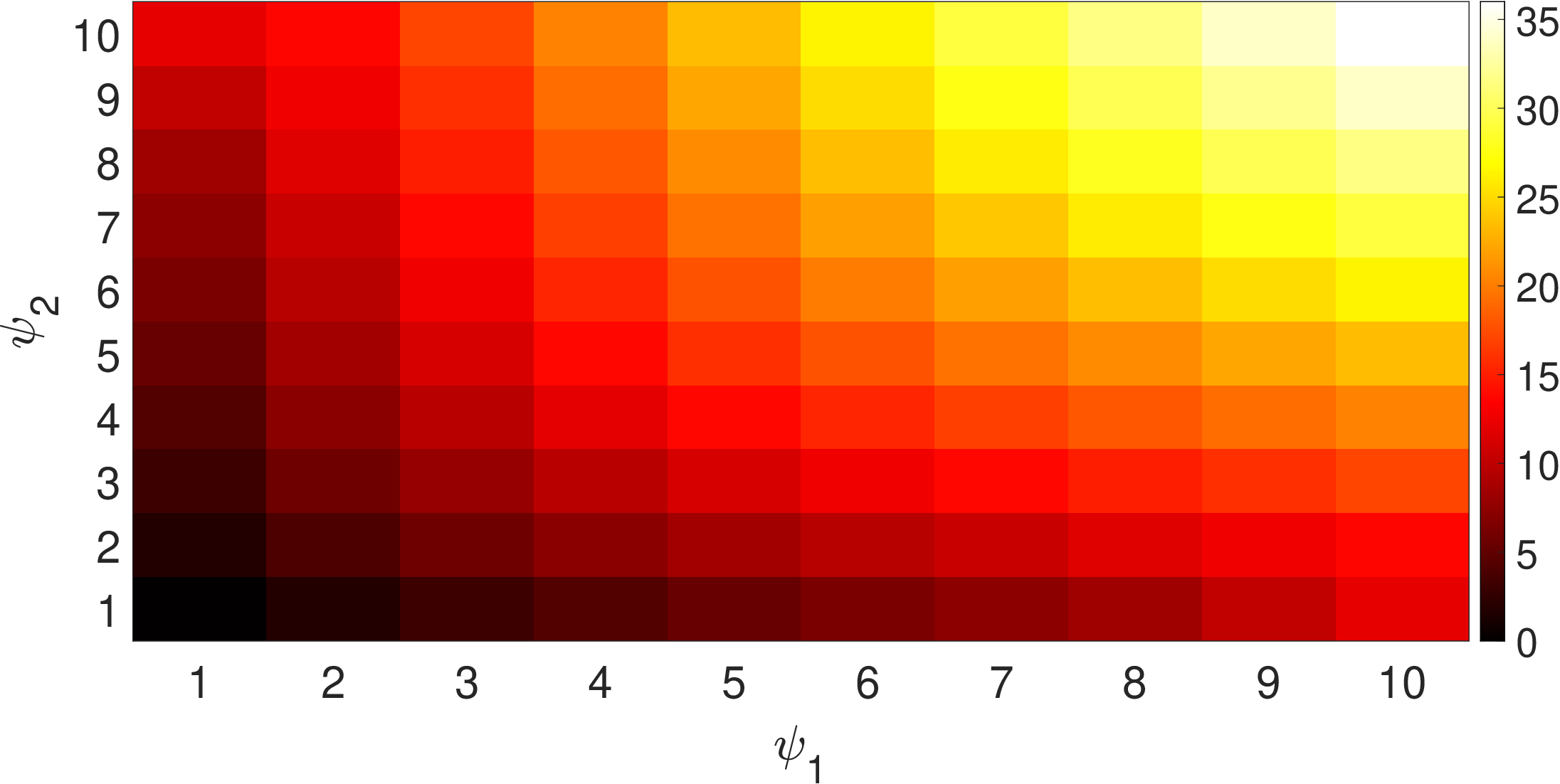}
        \label{fig:nb_ppt_b22}
    }
    \caption{Illustration of the public-private trade-off for Bargaining game ($\gamma_B$) against different values of $\psi_{n}(=r_{n}a_{n})$ for $N=2$ for $\textbf{b}=[1, 1]$, $\textbf{b}=[1, 2]$, and $\textbf{b}=[2, 2]$ in (a), (b), and (c), respectively.
    }
    \label{fig:NB_PPT}
\end{figure*}

Fig. \ref{fig:NB_BoB} illustrates the variation of the benefit of bargaining, i.e., $\beta$, for the Nash bargaining game against different values of $\psi_{n}(=r_{n}a_{n})$, where $n \in \{1, 2\}$, for $\textbf{b}=[1, 1]$, $\textbf{b}=[1, 2]$, and $\textbf{b}=[2, 2]$ in Fig. \ref{fig:bob_b11}, \ref{fig:bob_b12}, and \ref{fig:bob_b22}, respectively. 
We observe from Fig. \ref{fig:bob_b11} and \ref{fig:bob_b12} that $\beta=1$ when the CPs are comparable, i.e., when $\psi_{1} \approx \psi_{2}$. However, when CPs are non-comparable, i.e., either $\psi_{1} >> \psi_{2}$ or $\psi_{2} >> \psi_{1}$, the bargaining game results in higher public investment, and hence, $\beta > 1$. This follows from Theorem \ref{thm:NBS2}. Further, we observe from Fig. \ref{fig:bob_b11} that $\beta$ is symmetrical for $\psi_{1}>>\psi_{2}$ and $\psi_{2}>>\psi_{1}$ for $\textbf{b}=[1, 1]$. However, in the case of $\textbf{b}=[1, 2]$, $\beta$ is higher for $\psi_{1}>>\psi_{2}$ than $\psi_{2}>>\psi_{1}$ (Refer Fig. \ref{fig:bob_b12}). 
Although CP 2 dominates the total public investment as $\psi_{2} >> \psi_{1}$, it has an incentive for private investment since $\textbf{b}=[1, 2]$. Thus, both $Q_{B}^{\ast}$ and $Q_{C}^{\ast}$ reduces and in turn $\beta$ reduces. 
Similar to $\textbf{b}=[1, 2]$, $\beta$ reduces in the case of $\textbf{b}=[2, 2]$ as both CPs have an incentive for private investment.
Fig. \ref{fig:NB_PPT} illustrates the variation of the public-private trade-off for the Nash bargaining game, i.e., $\gamma_{B}$, against different values of $\psi_{n}(=r_{n}a_{n})$, where $n \in \{1, 2\}$, for $\textbf{b}=[1, 1]$, $\textbf{b}=[1, 2]$, and $\textbf{b}=[2, 2]$ in Fig. \ref{fig:nb_ppt_b11}, \ref{fig:nb_ppt_b12}, and \ref{fig:nb_ppt_b22}, respectively.
When the CPs are comparable, i.e., $\psi_{1} \approx \psi_{2}$, since $\beta=1$, $\gamma_{B}=\gamma_{C}$. $Q_{B}^{\ast}$ increases as $\psi_{n}$ increases for all $n=\{1, 2\}$. Consequently, the optimum private investment reduces, and $\gamma_{B}$ increases. Unlike $\textbf{b}=[1, 1]$, there is an incentive for CP(s) for private investment in $\textbf{b}=[1, 2]$ and $\textbf{b}=[2, 2]$. Thus, $\gamma_{B}$ is least for $\textbf{b}=[2, 2]$ and highest for $\textbf{b}=[1, 1]$. (Observe the range of $\gamma_{B}$ in the sidebar in Fig. \ref{fig:NB_PPT}.) 

\begin{figure*}
    \centering
    \subfigure[]
    {
        \includegraphics[height=1.6in,width=2in]{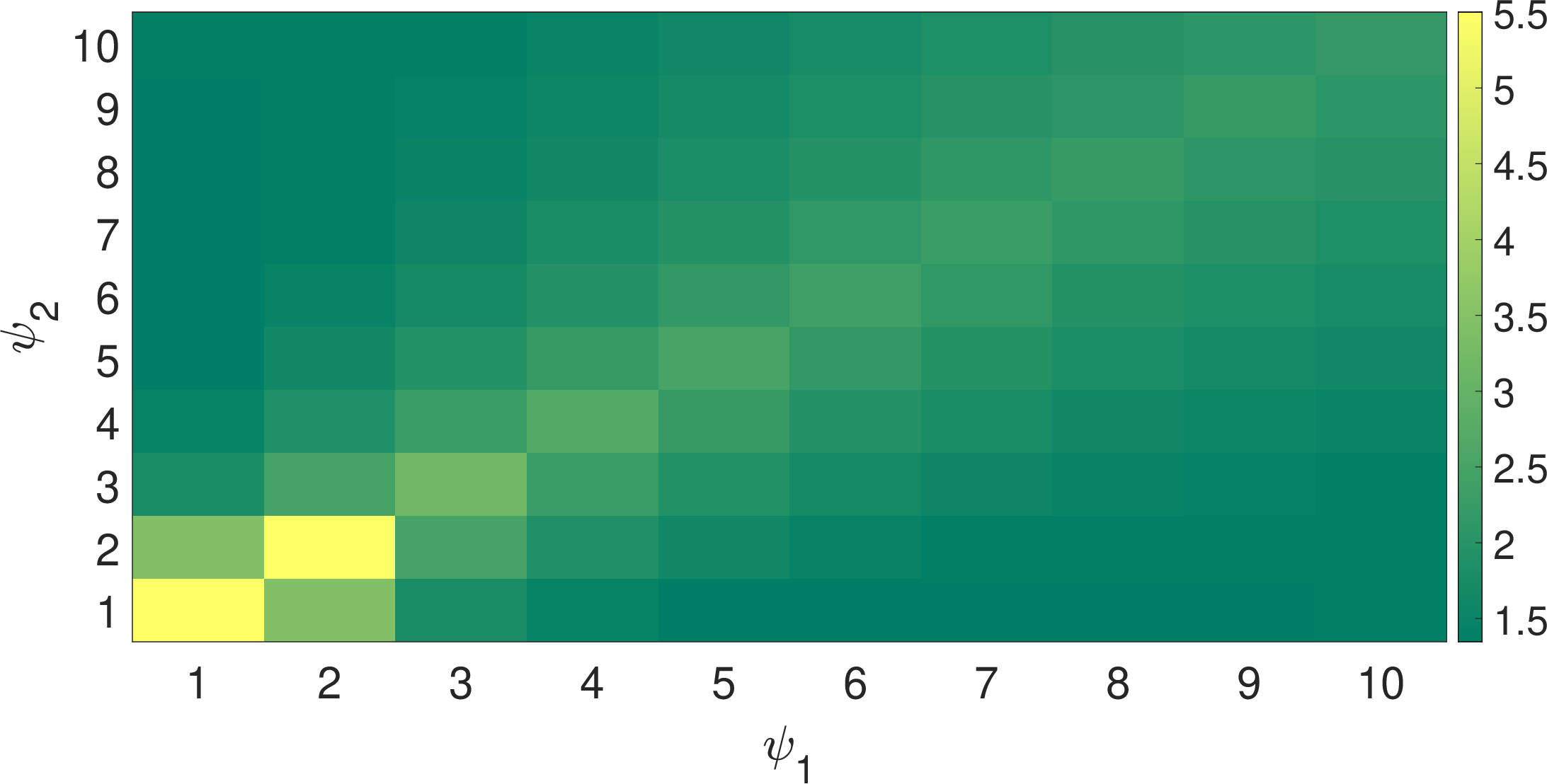}
        \label{fig:nb_ne_b11}
    }
    \subfigure[]
    {
        \includegraphics[height=1.6in,width=2in]{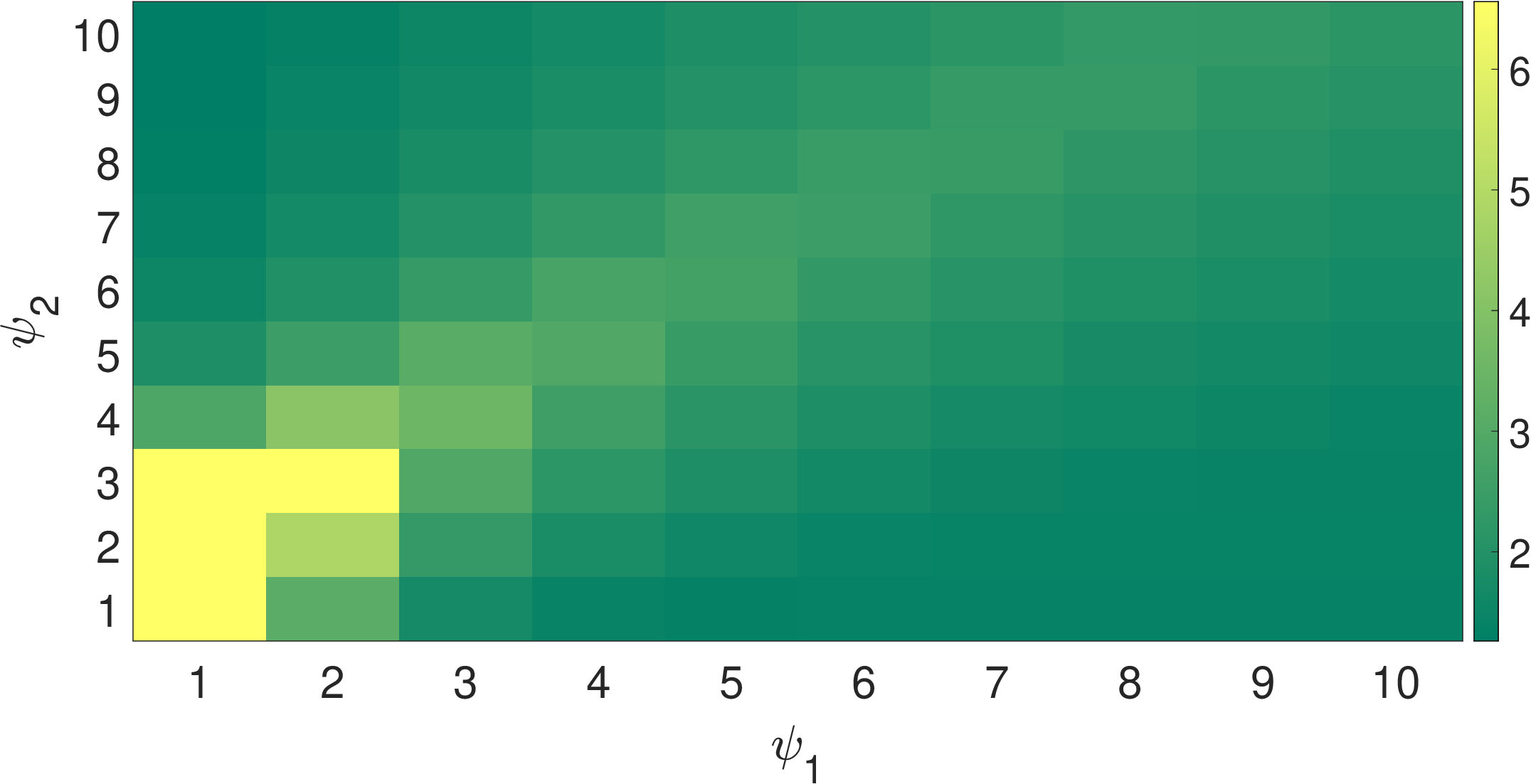}
        \label{fig:nb_ne_b12}
    }
    \subfigure[]
    {
        \includegraphics[height=1.6in,width=2in]{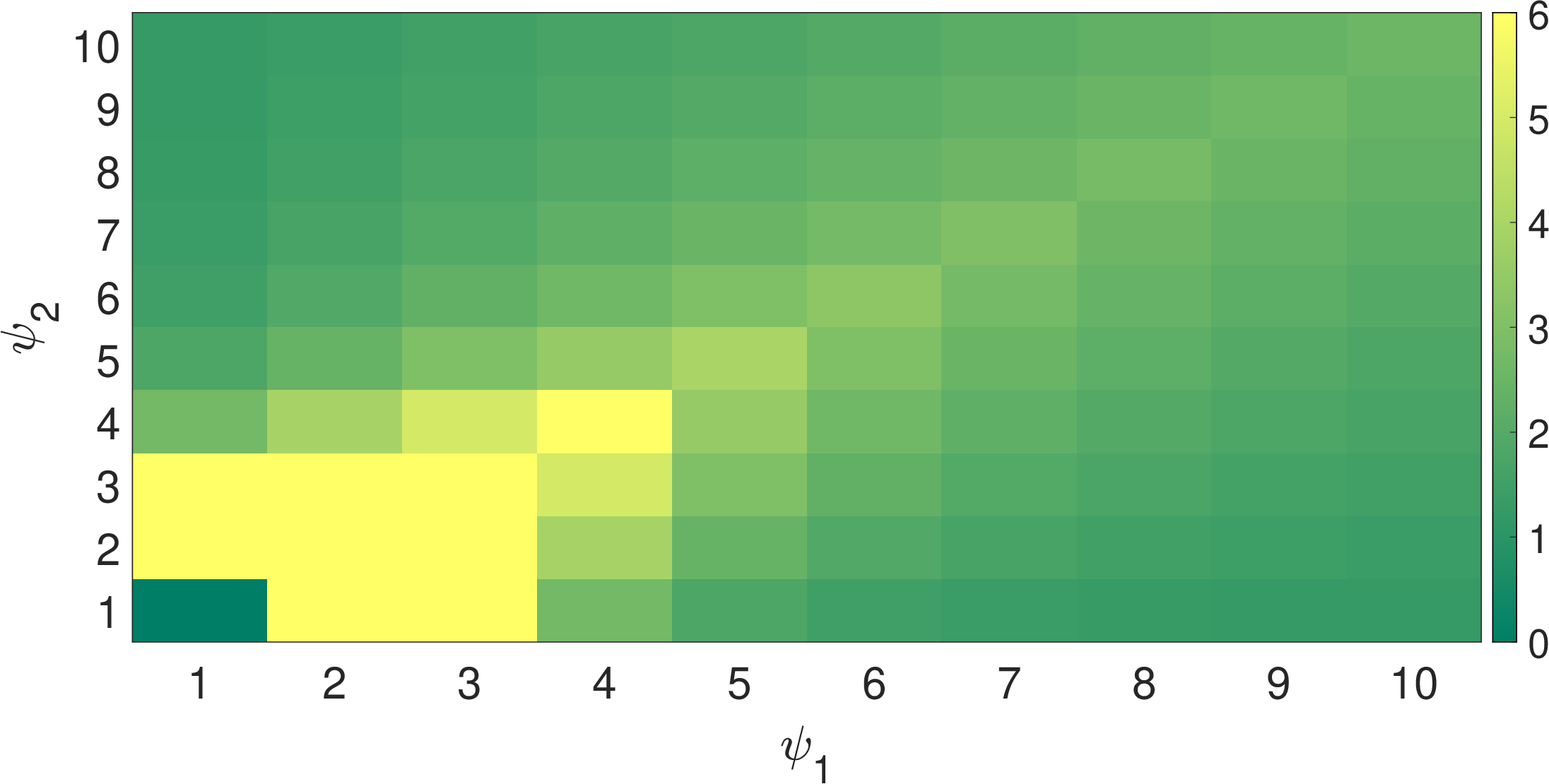}
        \label{fig:nb_ne_b22}
    }
    \caption{Illustration of the product of the price of anarchy and benefit of bargaining, i.e., $\alpha=\eta \times \beta$, against different values of $\psi_{n}(=r_{n}a_{n})$ for $N=2$ for $\textbf{b}=[1, 1]$, $\textbf{b}=[1, 2]$, and $\textbf{b}=[2, 2]$ in (a), (b), and (c), respectively.
    }
    \label{fig:NB_NE}
\end{figure*}

Fig. \ref{fig:NB_NE} illustrates the variation of the product of the price of anarchy and benefit of bargaining, i.e., $\alpha=\eta \times \beta$, against the different values of $\psi_{n}(=r_{n}a_{n})$, where $n \in \{1, 2\}$, for $\textbf{b}=[1, 1]$, $\textbf{b}=[1, 2]$, and $\textbf{b}=[2, 2]$ in Fig. \ref{fig:nb_ne_b11}, \ref{fig:nb_ne_b12}, and \ref{fig:nb_ne_b22}, respectively. In the case of $\textbf{b}=[1, 1]$, $Q_{N}^{\ast}=0$ for the region $(\psi_{1} \leq 1.5, \psi_{2} \leq 1.5)$ (Refer Theorem \ref{thm:NC}). Thus, $\alpha$ is unbounded as observed in Fig. \ref{fig:nb_ne_b11}. Similarly, we observe in Fig. \ref{fig:nb_ne_b12} that $\alpha$ is unbounded in the region $(\psi_{1} \leq 1.5, \psi_{2} \leq 3)$ when $\textbf{b}=[1, 2]$. For $\textbf{b}=[2, 2]$, $Q_{N}^{\ast}=0$ for the region $(\psi_{1} \leq 3, \psi_{2} \leq 3)$. Further, $Q_{B}^{\ast}=0$ for $(\psi_{1} \leq 1, \psi_{2} \leq 1)$ as $\sum \limits_{n \in \mathcal{N}} \frac{\sqrt{1+2b_{n}^{2}r_{n}a_{n}}-1}{b_{n}^{2}} \leq 1$ (Refer Lemma \ref{lemma:NBS}). Thus, for $(\psi_{1} \leq 1, \psi_{2} \leq 1)$, we consider $\alpha=0$. However, for $(1 < \psi_{1} \leq 3, 1 < \psi_{2} \leq 3)$, $Q_{B}^{\ast}>0$, and hence, $\alpha$ is unbounded. (Observe Fig. \ref{fig:nb_ne_b22}.) 
As we move away from $(\psi_{1}=1, \psi_{2}=1)$ along the diagonal, $Q_{N}^{\ast}$ increases from $0$ as $\psi_{n}$ increases for all $n \in \{1, 2\}$ and CPs are comparable, i.e., $\psi_{1} \approx \psi_{2}$. Thus, $\alpha$ reduces as observed in Fig. \ref{fig:NB_NE}.
Also, observe from Fig. \ref{fig:NB_NE} that $\alpha > 1$ (except for $(\psi_{1} \leq 1, \psi_{2} \leq 1)$ in Fig. \ref{fig:nb_ne_b22} as already discussed). This implies that the bargaining game results in higher public investment than the non-cooperative game.
\section{Conclusions}\label{sec:Conclusions}
The quality of the ISP access infrastructure impacts the profit of a CP. Thus, a CP can boost its profit by making a public investment to improve the ISP access infrastructure. A CP can also invest in its private infrastructure to boost its profit. Thus, in this work, we focus on the trade-off between public and private investments of CPs for four interaction models among CPs---a centralized allocation, a cooperative game, a non-cooperative game, and a bargaining game. The key findings of this work are as follows. In the strategic cooperative game, numerical results suggest that the grand coalition of all the CPs is feasible for certain scenarios. In the non-cooperative game, unless some symmetry exists among CPs, at most one CP makes the public investment. However, all CPs make private investments. The bargaining game can obtain higher public investment than the non-cooperative game and the centralized model. Thus, we see that if the regulators have a policy for public investments by CPs that emulates a bargaining game, then the public investments can be substantial.

\appendix
\section{Appendix}
\subsection{Proof of Lemma \ref{lem:opti_pn}}
\begin{proof}
The partial derivative of $U$, given by (\ref{eq:utility}), with respect to $p_{n}$ is given by
\begin{equation}
    \frac{\partial U}{\partial p_{n}} = \frac{r_{n}a_{n}b_{n}}{2\sqrt{p_{n}}(1+Q+b_{n}\sqrt{p_{n}})} - 1. \label{eq:partial_pn}
\end{equation}
From (\ref{eq:partial_pn}), we obtain the following quadratic equation in $\sqrt{p_{n}}$ by setting $\frac{\partial U}{\partial p_{n}}=0$,
\begin{equation}
    b_{n}p_{n} + \sqrt{p_{n}}(1+Q) - \frac{r_{n}a_{n}b_{n}}{2} = 0 . \label{eq:quadratic}
\end{equation}
The roots of (\ref{eq:quadratic}) are given by
\begin{equation}
    \sqrt{p_{n}} = \frac{-(1+Q) \pm \sqrt{(1+Q)^{2}+2b_{n}^{2}r_{n}a_{n}}}{2b_{n}}. \nonumber
\end{equation}
Since $a_{n}$, $r_{n}$, and $b_{n}$ are positive for all $n$, we have $\sqrt{(1+Q)^{2}+2b_{n}^{2}r_{n}a_{n}}>(1+Q)$. Further, $p_{n}$ has to be non-negative for all $n$. This implies the only possible solution of (\ref{eq:quadratic}) is
\begin{equation}
    \sqrt{p_{n}} = \frac{\sqrt{(1+Q)^{2}+2b_{n}^{2}r_{n}a_{n}}-(1+Q)}{2b_{n}} . \label{eq:pn_opti}
\end{equation}
From our discussion above, it follows that $p_{n}>0$ for any value of $Q$. 

To see that the $p_{n}$ is a convex decreasing function of $Q$, we consider $t=\sqrt{p_{n}}$. The partial derivative of $t$ with respect to $Q$ is given by
\begin{equation}
    \frac{\partial t}{\partial Q} = \frac{1}{2b_{n}}\left(\frac{(1+Q)-\sqrt{(1+Q)^{2}+2b_{n}^{2}r_{n}a_{n}}}{\sqrt{(1+Q)^{2}+2b_{n}^{2}r_{n}a_{n}}}\right). \label{eq:partial_t}
\end{equation}
It is easy to see from (\ref{eq:partial_t}) that $t$ is a decreasing function of $Q$. The second order derivative of $t$ with respect to $Q$ is given by
\begin{equation}
    \frac{\partial^{2}t}{\partial Q^{2}} = \frac{r_{n}a_{n}b_{n}}{((1+Q)^{2}+2b_{n}^{2}r_{n}a_{n})^{3/2}} . \label{eq:SOder_t}
\end{equation}
From (\ref{eq:SOder_t}), $\frac{\partial^{2}t}{\partial Q^{2}}>0$ which implies that $t$ is a convex function of $Q$.
Since $p_{n}=t^{2}$, we have
\begin{equation}
    \frac{\partial p_{n}}{\partial Q} = 2t\frac{\partial t}{\partial Q} . \label{eq:pn_der}
\end{equation}
Since $t>0$ and $\frac{\partial t}{\partial Q}<0$, it follows that $\frac{\partial p_{n}}{\partial Q}<0$. This implies that the $p_{n}$ is a decreasing function of $Q$. 
The second order derivative of $p_{n}$ with respect to $Q$ is given by
\begin{equation}
    \frac{\partial^{2}p_{n}}{\partial Q^{2}} = 2 \left(\frac{\partial t}{\partial Q}\right)^{2} + 2t\frac{\partial^{2}t}{\partial Q^{2}} . \nonumber 
\end{equation}
Since $t>0$, $\frac{\partial^{2} t}{\partial Q^{2}}>0$, and $\left(\frac{\partial t}{\partial Q}\right)^{2}>0$ for all $Q$, we have $\frac{\partial^{2} p_{n}}{\partial Q^{2}}>0$. Therefore, $p_{n}(Q)$ is a convex decreasing function of $Q$.
This completes the proof.
\end{proof}
\subsection{Proof of Lemma \ref{lem:lemma_coop}}
\begin{proof}
    The partial derivatives of $U$, given in (\ref{eq:utility}), with respect to $Q$ is computed as
    \begin{align}
        \frac{\partial U}{\partial Q} &= \sum_{n=1}^{N} \frac{r_{n}a_{n}}{1+Q+b_{n}\sqrt{p_{n}}} - 1 , \nonumber
    \end{align}
    Note that $\frac{\partial U}{\partial p_{n}}$ is given in (\ref{eq:partial_pn}), and hence, using (\ref{eq:pn_opti}), we obtain
    \begin{equation}
        \frac{\partial U}{\partial Q} = \sum_{n=1}^{N} \frac{2r_{n}a_{n}}{\sqrt{(1+Q)^{2}+2b_{n}^{2}r_{n}a_{n}}+(1+Q)} - 1, \nonumber
    \end{equation}
    which can be simplified to 
    \begin{equation}
        \frac{\partial U}{\partial Q} = \sum_{n=1}^{N} \left(\frac{\sqrt{(1+Q)^{2}+2b_{n}^{2}r_{n}a_{n}} - (1+Q)}{b_{n}^{2}}\right) - 1 . \label{eq:U_derivative}
    \end{equation}
    Clearly, (\ref{eq:U_derivative}) is the same as (\ref{eq:lemma_coop}).
    If $\left.\frac{\partial U}{\partial Q}\right\rvert_{Q=0}<0$, it implies $Q^{\ast}_{C}=0$. From (\ref{eq:U_derivative}), we have
    \begin{equation}
       \left. \frac{\partial U}{\partial Q}\right \rvert_{Q=0} = \sum_{n=1}^{N} \frac{\sqrt{1+2b_{n}^{2}r_{n}a_{n}}-1}{b_{n}^{2}} - 1. \nonumber
    \end{equation}
    Therefore, if $\sum_{n=1}^{N}\frac{\sqrt{1+2b_{n}^{2}r_{n}a_{n}}}{b_{n}^{2}}>1 + \sum_{n=1}^{N} \frac{1}{b_{n}^{2}}$, $\left.\frac{\partial U}{\partial Q}\right\rvert_{Q=0}>0$ which implies $Q^{\ast}_{C}>0$. Similarly, if $Q^{\ast}_{C}>0$, we have $\left.\frac{\partial U}{\partial Q}\right\rvert_{Q=0}>0$ which implies $\sum_{n=1}^{N} \frac{\sqrt{1+2b_{n}^{2}r_{n}a_{n}}}{b_{n}^{2}} > 1 + \sum_{n=1}^{N} \frac{1}{b_{n}^{2}}$. This completes the proof.  
\end{proof}
\subsection{Proof of Theorem \ref{thm:coop_gamma}}
\begin{proof}
    From Lemma \ref{lem:lemma_coop}, we know that $Q_{C}^{\ast}$ is obtained by solving (\ref{eq:lemma_coop}). Further, $p_{n}^{\ast}(Q)$ is given by (\ref{eq:opti_pn}). Thus, we have
    \begin{equation}
        p_{n}^{\ast}(Q) = \frac{(1+Q)}{2b_{n}^{2}} + \frac{r_{n}a_{n}}{2} - \frac{(1+Q)\sqrt{(1+Q)^{2}+2b_{n}^{2}r_{n}a_{n}}}{2b_{n}^{2}} \nonumber 
    \end{equation}
    Using (\ref{eq:lemma_coop}), we obtain
    \begin{align}
        \sum_{n=1}^{N} p_{n}^{\ast}(Q_{C}^{\ast}) &= \sum_{n=1}^{N}\frac{(1+Q_{C}^{\ast})}{2b_{n}^{2}} + \sum_{n=1}^{N} \frac{r_{n}a_{n}}{2} -  \frac{(1+Q_{C}^{\ast})}{2}\left(1 + (1+Q_{C}^{\ast})\sum_{n=1}^{N} \frac{1}{b_{n}^{2}}\right) , \nonumber \\
        &= \sum_{n=1}^{N} \frac{r_{n}a_{n}}{2} - \frac{(1+Q_{C}^{\ast})}{2} . \label{eq:gamma_coop_pn}
    \end{align}
    Substituting (\ref{eq:gamma_coop_pn}) into (\ref{eq:gamma}), we obtain (\ref{eq:lemma_coop_gamma}). This completes the proof.
\end{proof}

\subsection{Proof of Lemma \ref{lem:coalition}}
\begin{proof}
    Let $\mathcal{I}=\{i\}$ for $i \in \mathcal{N}$ such that $i>1$. This implies $q_{n}=0$ for all $n \in \mathcal{N}$ such that $n \neq i$. Therefore, $Q_{-i}=0$. The derivative of $U_{n}$ with respect to $q_{n}$, denoted by $\frac{\partial U_{n}}{\partial q_{n}}$, is given by
\begin{equation}
    \frac{\partial U_{n}}{\partial q_{n}} = \frac{\sqrt{(1+Q)^{2}+2b_{n}^{2}r_{n}a_{n}}-(1+Q)}{b_{n}^{2}} - 1. \nonumber
\end{equation}
Then, the derivative of $U_{i}$ with respect to $q_{i}$ when $Q_{-i}=0$ is given by
\begin{equation}
    \left.\frac{\partial U_{i}}{\partial q_{i}}\right \rvert_{Q_{-i}=0} = \frac{\sqrt{(1+q_{i})^{2}+2b_{i}^{2}r_{i}a_{i}}-(1+q_{i})}{b_{i}^{2}} - 1. \nonumber
\end{equation}
Let $q_{i}^{\ast}$ denote the optimum value of $q_{i}$ that maximizes $U_{i}$ when $Q_{-i}=0$. We obtain $q_{i}^{\ast}$ by setting $\left.\frac{\partial U_{i}}{\partial q_{i}}\right \rvert_{Q_{-i}=0} = 0$ as
\begin{equation}
    1 + q_{i}^{\ast} = r_{i}a_{i} - \frac{b_{i}^{2}}{2} . \label{eq:coal_proof1}
\end{equation}
For all $n \in \mathcal{N}$ such that $n \neq i$, we have $q_{n}=0$ and $Q_{-n}=q_{i}$. Then, $\left. \frac{\partial U_{n}}{\partial q_{n}} \right \rvert_{q_{n}=0} > 0$ implies $$\frac{\sqrt{(1+Q_{-n})^{2}+2b_{n}^{2}r_{n}a_{n}}-(1+Q_{-n})}{b_{n}^{2}} > 1$$ which in turn implies 
\begin{equation}
(1+Q_{-n}) < r_{n}a_{n} - \frac{b_{n}^{2}}{2} . \label{eq:coal_proof2}
\end{equation}
For all $n<i$, we have $r_{n}a_{n}>r_{i}a_{i}$ and $b_{n} \leq b_{i}$. Thus, from (\ref{eq:coal_proof1}) and (\ref{eq:coal_proof2}), we have $(1+q_{i}^{\ast})<r_{n}a_{n} - \frac{b_{n}^{2}}{2}$ which implies $\left. \frac{\partial U_{n}}{\partial q_{n}} \right \rvert_{q_{n}=0} >0$. This in turn implies that the CP $n$, where $n<i$, can boost its utility by making a non-zero public investment. It can be easily verified that a non-zero public investment of CP $n$ will also boost the utility of CP $i$. Therefore, if $q_{i}>0$, $q_{n}>0$, where $i \in \mathcal{N}$ such that $i>1$ and $n \in \mathcal{N}$ such that $n<i$. This completes the proof.
\end{proof}

\subsection{Proof of Theorem \ref{thm:NC}}
\begin{proof}
 From (\ref{eq:Un_2}), we obtain $\frac{\partial U_{n}}{\partial q_{n}}$ as
    \begin{equation}
        \frac{\partial U_{n}}{\partial q_{n}} = \frac{2r_{n}a_{n}}{\sqrt{(1+Q)^{2}+2b_{n}^{2}r_{n}a_{n}}+(1+Q)} - 1. \label{eq:Un_der2}
    \end{equation}
    From (\ref{eq:Un_der2}), we have
    \begin{align}
        \left.\frac{\partial U_{n}}{\partial q_{n}}\right \rvert_{Q=0}\leq 0 \implies \frac{2r_{n}a_{n}}{\sqrt{1+2b_{n}^{2}r_{n}a_{n}}+1} \leq 1 , \nonumber \\
        \implies \sqrt{1+2b_{n}^{2}r_{n}a_{n}} + 1 \geq 2r_{n}a_{n} \implies r_{n}a_{n} - \frac{b_{n}^{2}}{2} \leq 1 . \nonumber
    \end{align}
    Therefore, if $r_{n}a_{n}-\frac{b_{n}^{2}}{2} \leq 1$, we have $\left. \frac{\partial U_{n}}{\partial q_{n}}\right \rvert_{Q=0} \leq 0$.
    If $\max \limits_{n \in \mathcal{M}} \left(r_{n}a_{n}-\frac{b_{n}^{2}}{2}\right) \leq 1$, we have $r_{n}a_{n} - \frac{b_{n}^{2}}{2} \leq 1$ for all $n \in \mathcal{N}$.
    This in turn implies $\left.\frac{\partial U_{n}}{\partial q_{n}}\right \rvert_{Q=0} \leq 0$ for all $n \in \mathcal{N}$. From (\ref{eq:Un_der2}), we have
    \begin{equation}
        \left.\frac{\partial U_{n}}{\partial q_{n}} \right \rvert_{\{q_{n}=0, Q_{-n}>0\}} < \left.\frac{\partial U_{n}}{\partial q_{n}}\right \rvert_{\{q_{n}=Q_{-n}=0\}} \leq 0. \label{eq:NC_case1}
    \end{equation}
    It can be concluded from (\ref{eq:NC_case1}) that $q_{n}^{\ast}=0$ for any value of $Q_{-n}$ for all $n$. 

    We now consider the case when $\max \limits_{n \in \mathcal{N}} \left(r_{n}a_{n} - \frac{b_{n}^{2}}{2}\right) > 1$. 
    Since $\max \limits_{n \in \mathcal{N}} \left(r_{n}a_{n} - \frac{b_{n}^{2}}{2}\right) > 1$, $\left. \frac{\partial U_{n}}{\partial q_{n}} \right \rvert_{Q=0} >1$ holds for at least one CP, and hence, some CPs have an incentive to invest in the public investment. 
    Let us define $1+Q_{N}^{\ast} = \max \limits_{n \in \mathcal{N}} \left(r_{n}a_{n} - \frac{b_{n}^{2}}{2} \right)$. Note that $Q_{N}^{\ast}>0$ follows from $\max \limits_{n \in \mathcal{N}} \left(r_{n}a_{n} - \frac{b_{n}^{2}}{2}\right) > 1$. Further, $\left. \frac{\partial U_{n}}{\partial q_{n}} \right \rvert_{Q=Q_{N}^{\ast}} = 0$ for all $n \in \mathcal{M}$, where $\mathcal{M}$ is defined in (\ref{eq:set_M}).
    Let $\{\hat{q}_{n}\}_{n \in \mathcal{N}}$ such that $\hat{q}_{n} \geq 0$ for all $n \in \mathcal{N}$ be any action profile of the CPs. Further, let $1+\mu = \sum_{n \in \mathcal{N}} \hat{q}_{N}$. We now consider the following cases. If $1+\mu > 1+Q_{N}^{\ast}$, $\left. \frac{\partial U_{n}}{\partial q_{n}}\right \rvert_{Q=\mu} < 0$ for all $n \in \mathcal{M}$. Thus, some CPs have an incentive to decrease their public investment. This implies that $1+\mu$ is not a Nash Equilibrium. Similarly, if $1+\mu < 1+Q_{N}^{\ast}$, $\left. \frac{\partial U_{n}}{\partial q_{n}}\right \rvert_{Q=\mu} > 0$ for all $n \in \mathcal{M}$. Therefore, some CPs have an incentive to increase their public investment. Thus, it is not a Nash Equilibrium. Lastly, if $1+\mu = 1+Q_{N}^{\ast}$, $\left. \frac{\partial U_{n}}{\partial q_{n}}\right \rvert_{Q=\mu} = 0$ for all $n \in \mathcal{M}$. It is a Nash Equilibrium if $\hat{q}_{n}=0$ for all $n \in \mathcal{N} \setminus \mathcal{M}$. This completes the proof. 
\end{proof}
\subsection{Proof of Theorem \ref{thm:PoA}}
\begin{proof}
   We first consider the case when $\max\limits_{n \in \mathcal{N}} \left(r_{n}a_{n} - \frac{b_{n}^{2}}{2}\right) \leq 1$. This implies $\left(r_{n}a_{n} - \frac{b_{n}^{2}}{2}\right) \leq 1$ for all $n \in \mathcal{N}$. From Theorem \ref{thm:NC}, this in turn implies that $Q_{N}^{\ast}=0$. It can be verified that $\frac{\sqrt{1+2b_{n}^{2}r_{n}a_{n}}-1}{b_{n}^{2}} \leq 1$ follows from $r_{n}a_{n} - \frac{b_{n}^{2}}{2} \leq 1$ for all $n \in \mathcal{N}$. This leads to two possibilities which are as follows.
   \begin{itemize}
       \item $\sum_{n=1}^{N} \frac{\sqrt{1+2b_{n}^{2}r_{n}a_{n}}-1}{b_{n}^{2}} \leq 1$: In this case, from Lemma \ref{lem:lemma_coop}, we have $Q_{C}^{\ast}=0$, and hence, from (\ref{eq:eta}), it is clear that $\eta$ is undefined.
       \item $\sum_{n=1}^{N} \frac{\sqrt{1+2b_{n}^{2}r_{n}a_{n}}-1}{b_{n}^{2}} > 1$: In this case, from Lemma \ref{lem:lemma_coop}, we have $Q_{C}^{\ast}>0$. However, since $Q_{N}^{\ast}=0$, from (\ref{eq:eta}), it follows that $\eta$ is unbounded.
   \end{itemize}
   Since $Q_{N}^{\ast}=0$ when $\max\limits_{n \in \mathcal{N}} \left(r_{n}a_{n} - \frac{b_{n}^{2}}{2}\right) \leq 1$, from (\ref{eq:gamma}), we have $\gamma_{N}=0$.

   Now we consider that $\max\limits_{n \in \mathcal{N}}\left(r_{n}a_{n} - \frac{b_{n}^{2}}{2}\right) > 1$. 
   From Theorem \ref{thm:NC}, $\left.\frac{\partial U_{m}}{\partial q_{m}}\right \rvert_{q_{m}=Q_{N}^{\ast}} = 0$ follows for each $m \in \mathcal{M}$, where $\mathcal{M}$ is as defined in (\ref{eq:set_M}). This implies 
   \begin{equation}
       \frac{\sqrt{(1+Q_{N}^{\ast})^{2}+2b_{m}^{2}r_{m}a_{m}}-(1+Q_{N}^{\ast})}{b_{m}^{2}}=1 ~\forall m \in \mathcal{M}. \label{eq:PoA_pf1}
   \end{equation}
   Using (\ref{eq:PoA_pf1}), we have
   \begin{align}
       \left. \frac{\partial U}{\partial Q}\right \rvert_{Q=Q_{N}^{\ast}} &= \sum_{n \in \mathcal{N} \setminus \mathcal{M}} \frac{\sqrt{(1+Q_{N}^{\ast})^{2}+2b_{n}^{2}r_{n}a_{n}}-(1+Q_{N}^{\ast})}{b_{n}^{2}} + \sum_{n \in \mathcal{M}} \frac{\sqrt{(1+Q_{N}^{\ast})^{2}+2b_{n}^{2}r_{n}a_{n}}-(1+Q_{N}^{\ast})}{b_{n}^{2}} - 1, \nonumber
   \end{align}
    Thus, $\left. \frac{\partial U}{\partial Q}\right \rvert_{Q=Q_{N}^{\ast}}>0$. This implies $Q_{C}^{\ast}>Q_{N}^{\ast}$ as $\left. \frac{\partial U}{\partial Q}\right \rvert_{Q=Q_{C}^{\ast}} = 0$. This follows from the concavity of the utility function. Therefore, from (\ref{eq:eta}), $\eta>1$. Using (\ref{eq:opti_pn}), $p_{n}^{\ast}(Q_{N}^{\ast})$ for all $n \in \mathcal{N} \setminus \mathcal{M}$ is given by 
    \begin{equation}
        p_{n}^{\ast}(Q_{N}^{\ast}) = \left(\frac{\sqrt{(1+Q_{N}^{\ast})^{2}+2b_{n}^{2}r_{n}a_{n}}-(1+Q_{N}^{\ast})}{2b_{n}}\right)^{2}. \nonumber
    \end{equation}
    For all $n \in \mathcal{M}$, since $1+Q_{N}^{\ast} = r_{n}a_{n} - \frac{b_{n}^{2}}{2}$, we have
    \begin{equation}
    p_{n}^{\ast}(Q_{N}^{\ast}) = \frac{b_{n}^{2}}{4} . \nonumber
    \end{equation}
    Therefore, $\gamma_{N}$ is given by
    \begin{equation}
        \gamma_{N} = \frac{Q_{N}^{\ast}}{\sum \limits_{n \in \mathcal{M}} \frac{b_{n}^{2}}{4} + \sum \limits_{n \in \mathcal{N} \setminus \mathcal{M}} \left(\frac{\sqrt{(1+Q_{N}^{\ast})^{2}+2b_{n}^{2}r_{n}a_{n}}-(1+Q_{N}^{\ast})}{2b_{n}}\right)^{2}} . \label{eq:gamma_n2}
    \end{equation}
    Clearly, (\ref{eq:gamma_n}) follows from (\ref{eq:gamma_n2}) for the special case of $|\mathcal{M}|=|\mathcal{N}|$. This completes the proof.
\end{proof}
\subsection{Proof of Lemma \ref{lemma:NBS}}
\begin{proof}
    If $\sum \limits_{n \in \mathcal{N}} \frac{\sqrt{1+2b_{n}^{2}r_{n}a_{n}}-1}{b_{n}^{2}} > 1$, from Lemma \ref{lem:lemma_coop}, we have $Q_{C}^{\ast}>0$ which implies $U(Q_{C}^{\ast})>U(0) = \sum_{n=1}^{N} U_{n}^{\ast,D}$, where $U(Q_{C}^{\ast})$ and $U_{n}^{\ast, D}$ are given in (\ref{eq:utility}) and (\ref{eq:Un_D}), respectively. We can then determine some $\hat{q}_{1}, \cdots, \hat{q}_{N} \geq 0$ which satisfies the following set of equations
    \begin{align}
        \sum_{n=1}^{N} \hat{q}_{n} = Q_{C}^{\ast} , \nonumber \\
        U_{n}(Q_{C}^{\ast}) \geq U_{n}^{\ast, D}~\forall n . \nonumber
    \end{align}
    This implies $U_{1}(Q_{C}^{\ast}), \cdots, U_{N}(Q_{C}^{\ast}) \in \mathcal{U}$, and hence, we have a non-zero solution for (\ref{eq:NBS}).

    Let us consider that a non-zero solution exists for (\ref{eq:NBS}). Then, we have $\hat{q}_{1}, \cdots, \hat{q}_{N} \geq 0$ such that $U_{n}(\sum_{n=1}^{N} \hat{q}_{n})>U_{n}^{\ast, D}$ for all $n$. This implies $\sum_{n=1}^{N}U_{n}(\sum_{n=1}^{N} \hat{q}_{n}) > \sum_{n=1}^{N} U_{n}^{\ast, D}=U(0)$. Thus, $Q_{C}^{\ast}>0$ which in turn implies $\sum \limits_{n \in \mathcal{N}} \frac{\sqrt{1+2b_{n}^{2}r_{n}a_{n}}-1}{b_{n}^{2}} > 1$. This completes the proof.
\end{proof}
\subsection{Proof of Theorem \ref{thm:NBS}}
\begin{proof}
    Let us consider that there exist two optimizers of (\ref{eq:NBS}) which are denoted by $\hat{\textbf{U}}^{B}=(\hat{U}_{1}^{B}, \hat{U}_{2}^{B}, \cdots, \hat{U}_{N}^{B})$ and $\Tilde{\textbf{U}}^{B}=(\Tilde{U}_{1}^{B}, \Tilde{U}_{2}^{B}, \cdots, \Tilde{U}_{N}^{B})$. Thus, there exist $(\hat{q}_{1}, \hat{q}_{2}, \cdots, \hat{q}_{N})$ and $(\Tilde{q}_{1}, \Tilde{q}_{2}, \cdots, \Tilde{q}_{N})$ such that
    \begin{equation}
        \hat{U}_{n}^{B} = U_{n}(\hat{Q}) \text{ and } \Tilde{U}_{n}^{B} = U_{n}(\Tilde{Q}) ~\forall n, \label{eq:optimizers}
    \end{equation}
    where,
    \begin{equation}
        \hat{Q} = \sum_{n=1}^{N} \hat{q}_{n} \text{ and } \Tilde{Q} = \sum_{n=1}^{N} \Tilde{q}_{n} . \nonumber
    \end{equation}
    Further, $U_{n}(Q)$ and $f_{n}(Q)$ are as defined in (\ref{eq:Un_2}) and (\ref{eq:fnQ}), respectively.
    Note that $U_{n}(\hat{Q}) \geq U_{n}^{\ast, D}$ and $U_{n}(\Tilde{Q}) \geq U_{n}^{\ast, D}$ for all $n \in \mathcal{N}$.
    The first and second-order derivative of $f_{n}(Q)$ with respect to $Q$, denoted by $f_{n}'(Q)$ and $f_{n}''(Q)$, respectively, are given as
    \begin{align}
        f_{n}'(Q) &= \frac{1}{\sqrt{(1+Q)^{2}+2b_{n}^{2}r_{n}a_{n}}} , \nonumber \\
        f_{n}''(Q) &= \frac{-(1+Q)}{((1+Q)^{2}+2b_{n}^{2}r_{n}a_{n})^{3/2}} . \label{eq:f_der}
    \end{align}
    Since $Q \geq 0$, we have $f_{n}'(Q)>0$ and $f_{n}''(Q)<0$ for all $Q$ which implies that $f_{n}(Q)$ is a concave increasing function of $Q$. From Lemma \ref{lem:opti_pn}, we know that $p_{n}^{\ast}(Q)$ is a convex decreasing function of $Q$. Therefore, $U_{n}(Q)$, as given in (\ref{eq:Un_2}), is a concave function of $Q$.
    
    We now define $\Bar{U}_{n}^{B}$ as 
    \begin{align}
        \Bar{U}_{n}^{B} &= \Bar{U}_{n}(Q) = U_{n}\left(\frac{\hat{Q}+\Tilde{Q}}{2}\right) , \nonumber \\
                       &= r_{n}a_{n}f_{n}\left(\frac{\hat{Q}+\Tilde{Q}}{2}\right) - p_{n}^{\ast}\left(\frac{\hat{Q}+\Tilde{Q}}{2}\right) - \left(\frac{\hat{q}_{n}+\Tilde{q}_{n}}{2}\right) . \label{eq:bar_un}
    \end{align}
    Since $f_{n}(Q)$ and $p_{n}^{\ast}(Q)$ are concave and convex functions of $Q$, respectively, from (\ref{eq:optimizers})-(\ref{eq:bar_un}) and Jensen's inequality, we have for all $n \in \mathcal{N}$
    \begin{equation}
        (\Bar{U}_{n}(Q) - U_{n}^{\ast,D}) \geq  \frac{(\hat{U}_{n}(Q) - U_{n}^{\ast,D})+(\Tilde{U}_{n}(Q) - U_{n}^{\ast,D})}{2} . \label{eq:bar_Un_ineq}
    \end{equation}
    Using arithmetic and geometric mean inequality, from (\ref{eq:bar_Un_ineq}), we have
    \begin{equation}
        \prod_{n \in \mathcal{N}} \frac{(\hat{U}_{n}(Q) - U_{n}^{\ast,D})+(\Tilde{U}_{n}(Q) - U_{n}^{\ast,D})}{2} \geq \prod_{n \in \mathcal{N}} (U_{n}(\hat{Q})-U_{n}^{\ast,D}) = \prod_{n \in \mathcal{N}} (U_{n}(\Tilde{Q}) - U_{n}^{\ast,D}) . \label{eq:AM-GM}
    \end{equation}
    From (\ref{eq:bar_Un_ineq}) and (\ref{eq:AM-GM}), we have for all $n \in \mathcal{N}$
    \begin{equation}
        \prod_{n=1}^{N} (\Bar{U}_{n}^{B} - U_{n}^{\ast,D}) > \prod_{n=1}^{N} (\hat{U}_{n}^{B} - U_{n}^{\ast,D}) = \prod_{n=1}^{N} (\Tilde{U}_{n}^{B} - U_{n}^{\ast,D}). \nonumber
    \end{equation}
    Since $(\Bar{U}_{1}^{B}, \Bar{U}_{2}^{B}, \cdots, \Bar{U}_{N}^{B}) \in \mathcal{U}$, there is a contradiction. This completes the proof.
\end{proof}
\subsection{Proof of Theorem \ref{thm:NBS2}}
\begin{proof} 
    The first part of the Theorem \ref{thm:NBS2} follows from Lemma \ref{lemma:NBS} and proof of Theorem $3$ in \citet{Kalvit2019}. Let us now prove the second part.
    Using (\ref{eq:Un_2}), the objective function for the Nash Bargaining problem as defined in (\ref{eq:NBS}) is given by
    \begin{equation}
        \zeta(Q) = \prod_{n=1}^{N} (r_{n}a_{n}f_{n}(Q) - p_{n}^{\ast}(Q) - q_{n}) - U_{n}^{\ast, D} . \nonumber
    \end{equation}
    The first order derivative of $\zeta(Q)$ with respect to $q_{m}$ is given by
    \begin{align}
        \frac{\partial \zeta(Q)}{\partial q_{m}} = \zeta(Q)\frac{r_{m}a_{m}f_{m}'(Q)-p_{m}^{\ast '}(Q)-1}{r_{m}a_{m}f_{m}(Q)-p_{m}^{\ast}(Q)-q_{m} - U_{m}^{\ast, D}} + \sum_{n=1, n \neq m}^{N} \zeta(Q) \frac{r_{n}a_{n}f_{n}'(Q)-p_{n}^{\ast '}(Q)}{r_{n}a_{n}f_{n}(Q)-p_{n}^{\ast}(Q)-q_{n} - U_{n}^{\ast,D}} . \nonumber
    \end{align}
    Setting $\frac{\partial \zeta(Q)}{\partial q_{m}}=0$ and simplifying further, we obtain  
 	\begin{align}
 	\sum_{n=1}^{N} \frac{r_{n}a_{n}f_{n}'(Q)-p_{n}^{\ast '}(Q)}{r_{n}a_{n}f_{n}(Q)-p_{n}^{\ast}(Q)-q_{n} - U_{n}^{\ast, D}} =
  \frac{1}{r_{m}a_{m}f_{m}(Q) - p_{m}^{\ast}(Q) - q_{m} - U_{m}^{\ast, D}}~\forall m \in \mathcal{N}. \nonumber
 	\end{align}
    This implies 
    \begin{equation}
        r_{n}a_{n}f_{n}(Q) - p_{n}^{\ast}(Q) - q_{n} - U_{n}^{\ast, D} = \text{constant} \nonumber
    \end{equation}
    for all $n \in \mathcal{N}$, and
    \begin{equation}
        \sum_{n=1}^{N} r_{n}a_{n}f_{n}'(Q) - p_{n}^{\ast '}(Q) = 1 . \label{eq:NBS2_cond}
    \end{equation}
    Substituting (\ref{eq:partial_t}), (\ref{eq:pn_der}), and (\ref{eq:f_der}) into (\ref{eq:NBS2_cond}), we have
    \begin{align}
        \sum_{n=1}^{N} \frac{(1\!+\!Q)^{2}\!+\!2b_{n}^{2}r_{n}a_{n} \!-\! (1\!+\!Q)\sqrt{(1\!+\!Q)^{2}\!+\!2b_{n}^{2}r_{n}a_{n}}}{\sqrt{(1\!+\!Q)^{2}\!+\!2b_{n}^{2}r_{n}a_{n}}} &= 1, \nonumber \\
        \sum_{n=1}^{N} \frac{\sqrt{(1+Q)^{2}+2a_{n}} - (1+Q)}{b_{n}^{2}} &= 1 , \nonumber
    \end{align}
    which is the same as (\ref{eq:lemma_coop}). Therefore, $Q_{C}^{\ast}=Q_{B}^{\ast}$. This completes the proof.
\end{proof}

\bibliographystyle{cas-model2-names}

\bibliography{Pranay_PEVA_Arxiv}

%
%
%

\end{document}